\documentclass[twocolumn]{aastex631}

\usepackage{multirow}

\begin{document}

\title{Morphology classification for galaxies in the Kilo Degree Survey using a label-efficient self-supervised learning framework}
%\title{\textcolor{red}{Efficient Galaxy Morphology Classification in the Kilo Degree Survey via Self-Supervised Learning}}

\author{Xu Huang}
\affiliation{Institute for Astrophysics, School of Physics, Zhengzhou University, Zhengzhou 450001, China}

\author{Rui Li}
\affiliation{Institute for Astrophysics, School of Physics, Zhengzhou University, Zhengzhou 450001, China}

\author{Liang Gao}
\affiliation{Institute for Astrophysics, School of Physics, Zhengzhou University, Zhengzhou 450001, China}
\affiliation{School of Physics and Astronomy, Beijing Normal University, Beijing 100875, China}

\author{Liqing Chen}
\affiliation{Institute for Astrophysics, School of Physics, Zhengzhou University, Zhengzhou 450001, China}

\author{Hui Li}
\affiliation{Institute for Astrophysics, School of Physics, Zhengzhou University, Zhengzhou 450001, China}

\author{Huijun Mu}
\affiliation{Institute for Astrophysics, School of Physics, Zhengzhou University, Zhengzhou 450001, China}

\author{Hao Su}
\affiliation{Department of Physics ``E. Pancini'', University Federico II, Via Cinthia 6, 80126 Napoli, Italy}

\author{Fucheng Zhong}
\affiliation{School of Physics and Astronomy, Sun Yat-sen University, Zhuhai Campus, 2 Daxue Road, Xiangzhou District, Zhuhai, China}

\author{Zhenping Yi}
\affiliation{School of Mechanical, Electrical and Information Engineering, Shandong University, China}

\author{Xiaoyue Cao}
\affiliation{Institute for Astrophysics, School of Physics, Zhengzhou University, Zhengzhou 450001, China}

\author{Ran Li}
\affiliation{School of Physics and Astronomy, Beijing Normal University, Beijing 100875, China}

\author{Haicheng Feng}
\affiliation{Yunnan Observatories, Chinese Academy of Sciences, Kunming 650216, Yunnan, China}

\author{Nicola N. Napolitano}
\affiliation{Department of Physics ``E. Pancini'', University Federico II, Via Cinthia 6, 80126 Napoli, Italy}

\author{Yue Dong}
\affiliation{School of Mathematics and Physics, Xi'an Jiaotong-Liverpool University, Suzhou 215123, China}

\author{Yida Deng}
\affiliation{Institute for Astrophysics, School of Physics, Zhengzhou University, Zhengzhou 450001, China}

\author{Sihan Li}
\affiliation{Institute for Astrophysics, School of Physics, Zhengzhou University, Zhengzhou 450001, China}

\author{Kang Jiao}
\affiliation{Institute for Astrophysics, School of Physics, Zhengzhou University, Zhengzhou 450001, China}

\begin{abstract}
Galaxy morphology classification is fundamental to understanding galaxy formation and evolution. The advent of large-scale sky surveys has produced an unprecedented volume of galaxy images, making traditional manual classification impractical. Although supervised deep learning can achieve high accuracy, it requires large labeled datasets that are time-consuming to construct. In contrast, unsupervised methods often show limited classification performance.
To address this limitation, we propose a label-efficient self-supervised learning framework for galaxy morphology classification. 
Our method first learns robust morphological representations from 305,583 unlabeled KiDS galaxy images through contrastive learning, and then trains a classifier using only 5,000 human-labeled images. 
The classifier separates galaxies into five categories: elliptical, spiral, lenticular-disk, irregular, and ``other.'' 
Using a ResNet-50 model with a crop size of $64\times64$ pixels,
our approach achieves an overall test accuracy of up to $91.0\%$ ($90.5\%\pm0.2\%$ on average) on the human-classified catalog. The corresponding F1 scores for elliptical, spiral, irregular, lenticular-disk, and ``other'' 
galaxies are 0.96, 0.86, 0.86, 0.95, and 0.92, respectively.
We apply this pipeline to the Kilo-Degree Survey Data Release 5 and produce a publicly available morphology catalog of 310,583 galaxies. 
This is the first morphology catalog for KiDS galaxies and provides a valuable resource for future studies of galaxy evolution. 
Our results show that self-supervised learning can substantially reduce the need for manual labels while maintaining high classification accuracy, making it a promising and scalable approach for automated galaxy morphology classification in the era of large-scale surveys.
\end{abstract}

% \keywords{\uat{Galaxy classification}{251} --- \uat{Self-supervised learning}{1736} --- \uat{History of astronomy}{1868} --- \uat{Interdisciplinary astronomy}{804}}

\section{Introduction}
Galaxies exhibit remarkable diversity in their structures and physical properties, ranging from massive ellipticals dominated by old stellar populations to spiral systems with active star formation to irregular dwarfs with chaotic appearances (e.g., \citealt{spiral1998, ellptical2006,  dwarf2009}). Galaxy morphology, which characterizes these structural features and spatial arrangements of stars, gas, and dust, serves as a fundamental tool for understanding the physical processes driving galaxy formation and evolution \citep{GalaxyZoo1, conselice, Krywult}. This morphological diversity makes classification a critical step in unraveling the complex evolutionary pathways of galaxies, providing insights into both the hierarchical assembly paradigm of structure formation and the environmental factors that shape galaxy properties across cosmic time (e.g., \citealt{gini2004, Dressler1980}). Observational studies further show that galaxy morphology evolves strongly with redshift, with irregular and disk-dominated systems being far more common at high redshift, while the fraction of spheroidal, early-type galaxies increases toward the present day—reflecting the gradual transformation from star-forming disks into quenched elliptical systems over cosmic time (e.g., \citealt{Huertas2016, Huerytas2024, Conselice2020, Holwerda2021}).

   The significance of galaxy morphology has been recognized since Hubble introduced the `Hubble sequence' for classification, categorizing galaxies into elliptical, spiral, and irregular types \citep{hubble1926}. This pioneering work established the first classification framework based on visual morphological features and laid the foundation for subsequent methods. 
   % Later developments included parametric approaches such as the Sérsic index $n$ derived from the Sérsic model \citep{Sersic, Sen}; 
   Later developments introduced parametric profile-fitting methods. Among these, the Sérsic model \citep{Sersic} is one of the most commonly used, as it allows for the derivation of structural parameters such as the Sérsic index $n$. This parameter reflects the morphology of different galaxies; for example, n$>$2.5 typically indicates elliptical galaxies in an evolved stage (e.g., \citealt{Roy2018Sersic}).
   These approaches have become standard in quantitative morphology studies and are widely applied in both earlier works (e.g., \citealt{Sen, HSTfit2011}) and in recent large imaging surveys, including those based on JWST data \citep{JWST2025}, KiDS data \citep{KiDSfit2025} and Euclid data \citep{Euclid2025}.
   In parallel, non-parametric morphology descriptors provide an alternative way to characterize the light distribution of galaxies without assuming analytic profiles.
   Classical examples include the CAS (Concentration, Asymmetry, Smoothness) parameters \citep{CAS}, and the Gini coefficient \citep{gini, gini2004}, which capture structural features such as central compactness, degree of disturbance, and the distribution of flux among pixels. These indicators remain widely used in modern observational studies \citep{CAScite2024, CAScitejwst2023, neareast2024, gini2016}. These techniques all aim to quantify morphological characteristics through measurable features.
    
    With the advent of large-scale astronomical surveys, the volume of galaxy images has increased exponentially. The third generation of surveys, such as the Sloan Digital Sky Survey (SDSS; \citealt{SDSS}), Dark Energy Survey (DES; \citealt{DES}), and the Kilo-Degree Survey (KiDS; \citealt{KiDS2013}), has produced unprecedented amounts of data, while upcoming fourth-generation surveys, including Euclid \citep{Euclid}, the Large Synoptic Survey Telescope (LSST; \citealt{LSST}), and the Chinese Space Station Telescope (CSST; \citealt{CSST}) promise even greater yields. This data explosion creates both opportunities and challenges for galaxy morphology studies, demanding classification methods that are both highly accurate and computationally efficient.
    
    Traditional classification methods like manual inspection, while valuable, face significant limitations in the era of big data. The Galaxy Zoo project \citep{GalaxyZoo}, a citizen science initiative that engages volunteers to visually classify galaxies from surveys such as the Sloan Digital Sky Survey (SDSS; \citealt{SDSS}), has demonstrated the power of human classification. However, manual approaches are inherently limited by their labor-intensive nature, subjective biases, and impracticality for processing millions of images generated by modern surveys.
    
    On the other hand, parametric approaches, while foundational to galaxy morphological studies, also face considerable limitations when confronted with the scale of modern astronomical surveys. Obtaining the Sérsic index, for instance, requires fitting the Sérsic profile to individual galaxies—a computationally intensive process that becomes prohibitively challenging when applied to millions of objects \citep{GaLnet}. These fits are highly sensitive to image quality, background noise, and resolution, often leading to inconsistent results across different datasets or observational conditions \citep{sersic2005}. The parameter extraction process itself introduces a layer of model-dependent assumptions that may not uniformly apply across the full morphological diversity of galaxies, particularly for peculiar, interacting, or high-redshift systems where classical morphological frameworks break down \citep{CAS}. A fundamental limitation of Sérsic fitting lies in its assumption of smooth and symmetric light distributions, an assumption that fails to capture the clumpy and asymmetric structures characteristic of irregular galaxies \citep{gini2004, parmwrong2011, Sersic2019, Sericspiral2022}. 
    
    In contrast, non-parametric methods such as the CAS parameters and the Gini coefficient aim to avoid model-specific biases, but they are not without challenges. 
    The CAS parameters similarly suffer from strong sensitivity to image quality, including background noise, sky-subtraction residuals, and PSF variations, with asymmetry measurements particularly vulnerable to contamination and noise \citep{asym, CAS}.
    The Gini coefficient, while useful for quantifying light distribution, is insensitive to the spatial location of pixels; as a result, it cannot fully capture structural features such as multiple components or irregular morphologies \citep{gini08}. 
    Furthermore, these non-parametric methods often require careful tuning and preprocessing for each specific dataset, making them difficult to standardize across diverse survey conditions \citep{gini08, parmwrong2011}. These limitations become increasingly problematic as survey data volumes expand, creating a bottleneck in the classification pipeline and motivating the search for more scalable and robust classification methodologies.
    
    The rise of deep learning has revolutionized galaxy classification through convolutional neural networks (CNNs). Various CNN architectures have been applied successfully, including ResNet variants \citep{cnn1}, which achieved 95.21\% accuracy across five morphological types on a subset of Galaxy Zoo 2 (GZ2) data, and EfficientNet \citep{cnn2}, which attained 93.7\% accuracy on another curated GZ2 subset. Other notable approaches include hierarchical classification using Bayesian networks combined with CNNs \citep{Beys}, transfer learning with pretrained models \citep{pre}, and various CNN-based classifiers \citep{rotation, densenet1, densenet2, densenet3, twoway, imp, dense4_2025}. Despite these advances, supervised learning methods remain constrained by their dependence on large amounts of labeled data, which are often scarce in astronomical contexts.
    
    To address the limitations of labeled data dependency, researchers have explored unsupervised learning methods. These approaches include variational autoencoders such as AstroVaDEr \citep{coder}, convolutional autoencoders for feature extraction and image similarity retrieval \citep{autoencoder0, autoencoder}, and self-organized maps as implemented in the EGG system \citep{hybird}, which achieved 95\% accuracy in a two-class unsupervised classification task. 
    % \textcolor{red}{To boost the efficiency of unsupervised galaxy morphology classification, \cite{volcluster2025} introduced an encoding scheme built upon the ConvNeXt large model.
 However, unsupervised methods often lack a clear guidance signal, limiting their ability to learn representations that transfer effectively across diverse tasks.

This limitation motivates the development of methods that can maintain high classification accuracy with only a limited number of labeled images. 
Self-supervised learning (SSL) provides a promising solution by using large volumes of unlabeled data to learn useful and transferable representations. 
In SSL, models are first trained on pretext tasks constructed from the data themselves, without manual labels. 
The learned representations are then transferred to downstream tasks, where only a small labeled sample is required for supervised training or fine-tuning.
SSL has recently been adopted in several areas of astrophysics. 
It has been applied to radio source classification \citep{BaronPerez2025}, strong gravitational lens detection \citep{Yang2025}, and time-domain astronomy \citep{Zuo2025}. 
These applications demonstrate the ability of SSL to extract informative features from astronomical data while reducing reliance on large labeled datasets.
For galaxy morphology classification, SSL has shown similarly encouraging results. 
\cite{Hayat2021} showed that SSL can reach accuracy comparable to supervised learning while requiring far fewer labeled galaxy images. 
Other SSL-based approaches using vision transformers and convolutional networks achieved test accuracies of 94.7\%, 96.5\%, and 89.9\% on the three-class GZ2, SDSS-DR17, and four-class GZ DECaLS datasets, respectively \citep{selfsuper1}. 
MCL-Galaxy, a momentum contrastive learning method, obtained an accuracy of 90.12\% on a filtered subset of Galaxy Zoo 2 \citep{selfsuper2}.
\cite{2026ApJS..283...49L} investigated the effectiveness of masked-autoencoder-based SSL for astronomical image analysis on DESI data, focusing on downstream tasks including galaxy morphology classification, object detection, and photometric redshift estimation, and showed that the SSL framework outperformed supervised learning when trained with the same amount of data.

Beyond the SSL methods mentioned above, contrastive learning represents another widely used SSL paradigm that learns representations by bringing different augmented views of the same object closer while separating views of different objects.
In this paper, we propose a novel approach for galaxy classification using SimCLR \citep{simclr2020}, a self-supervised contrastive learning framework. Our method enables the model to learn robust features from unlabeled data; then we train a classifier on labeled data to enhance performance specifically for galaxy classification tasks. We apply this approach to classify a subset of galaxies from the Kilo-Degree Survey (KiDS) Data Release 5 into five morphological categories. Given the scale and depth of the KiDS survey, our method addresses both the accuracy requirements and computational challenges inherent in modern astronomical datasets, potentially offering a more efficient pathway for morphological classification in upcoming large-scale surveys.
    
    The paper is organized as follows. Section \ref{sec:data} introduces the KiDS DR5 dataset used in this study. Section \ref{sec:method} describes the architecture of our self-supervised model and the training process. Section \ref{sec:performance} provides a systematic evaluation of classification performance on the reserved test set. Section \ref{sec: ablation} analyzes the impact of different model configurations. Section \ref{sec:apply} demonstrates the application of the trained model to the KiDS DR5 dataset. 
    Section \ref{sec:discussion} compares our method with existing approaches and discusses broader implications.
    % Section \ref{sec:discussion} discusses the results and their implications, including the effect of training settings and relevance to future surveys.
    Section \ref{sec:conslusion} concludes the paper and outlines directions for future work.

%__________________________________________________________________


\section{Data} \label{sec:data}

% \section{Data}
Our study utilizes public data from the Kilo-Degree Survey (KiDS) Data Release 5 (KiDS DR5; \citealt{KiDS5}). KiDS is a wide-field imaging survey conducted by the European Southern Observatory (ESO), primarily designed for weak gravitational lensing studies. KiDS DR5 provides observations in four optical bands ($u,g,r,i$) covering 1350 $\rm deg^2$ of the sky. The survey achieves a median r-band seeing of $0.7^{\prime\prime}$, with a limiting AB magnitude of 25.2. Table \ref{tab:observation_propeties} summarizes the exposure times, limiting magnitudes, and seeing for all four bands. For our morphological analysis, we constructed $gri$ color-composite images. Specifically, we used the g-band (467 nm) for the blue channel, the r-band (617 nm) for the green channel, and the i-band (748 nm) for the red channel. These composites are designed to clearly reveal structural features while maintaining physical significance. The resulting images have a pixel scale of $0.20''$/pixel.

\begin{figure*}[htbp]
% \plotone{4sho2.pdf}
\centering
\includegraphics[width=0.95\textwidth]{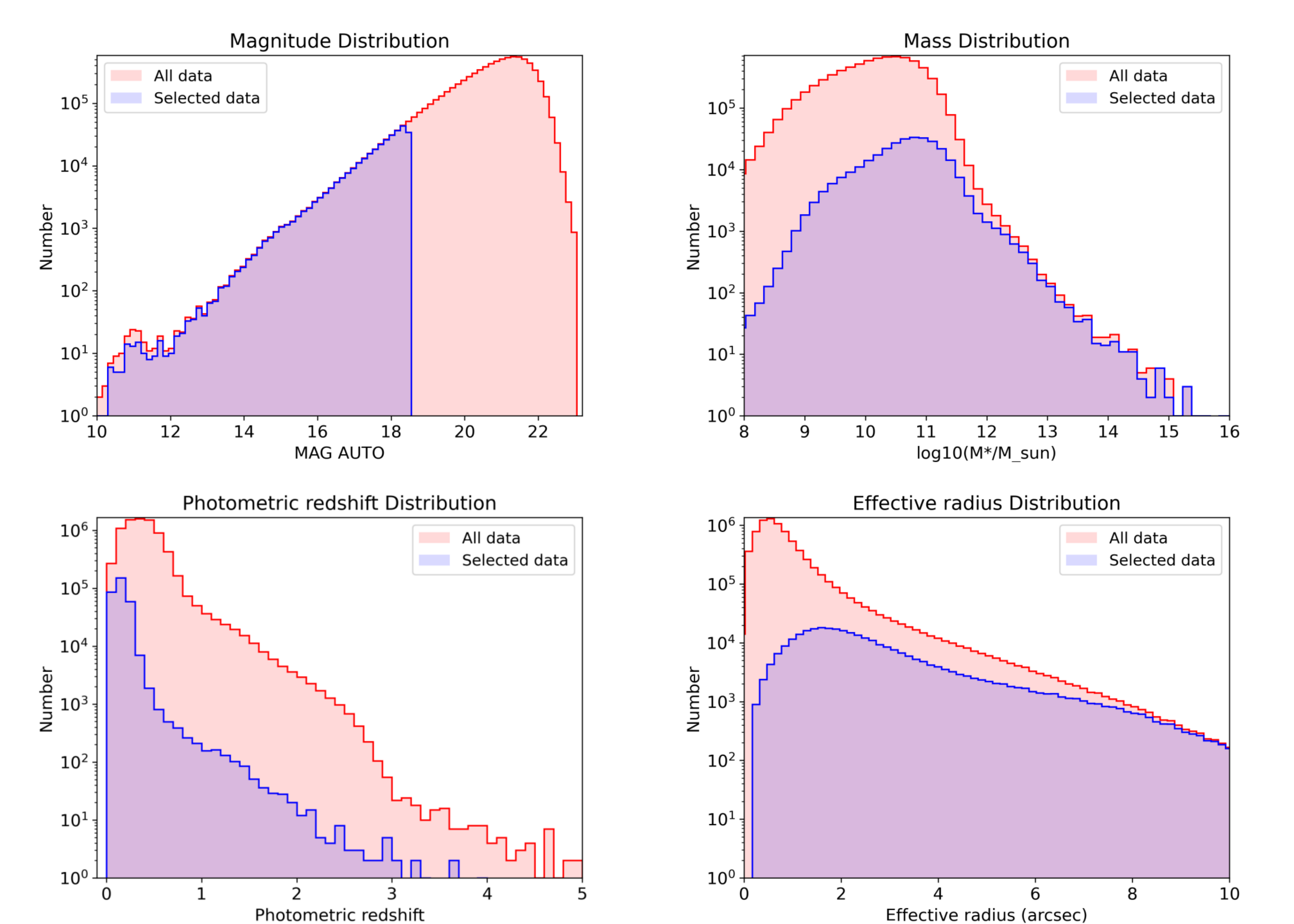}
    % \caption{Distributions of fundamental parameters for the 310,000-galaxy sample: magnitude, stellar mass, Sérsic index $n$ and effective radius. Note: The vertical axis for magnitude is logarithmic in counts; all other axes are linear.}
    \caption{Distributions of fundamental parameters ($r$-band magnitude, stellar mass, photometric redshift, and effective radius) for the selected sample of 310,583 galaxies compared to the full KiDS dataset (limited to S/N $> 35$). The $r$-band magnitudes are derived from SExtractor \citep{Sextractor} and stellar masses from CIGALE \citep{mass}, 
    while the effective radii are inferred by GaLNet \citep{GaLnet} and the photometric redshifts are estimated using the GaZNet \citep{GaZnet} machine learning tool.}
    % while the S\'ersic indices and effective radii are inferred by GaLNet \citep{GaLnet}. Note that the vertical axes are plotted on a logarithmic scale.}
    \label{fig:4}
\end{figure*}

The sample used in this study is drawn from the galaxy/star/QSO classification catalog of \cite{2025ApJS..279...26F}. In that work, the authors constructed deep neural networks to classify KiDS DR5 targets into stars, galaxies, and QSOs. The networks achieve an overall accuracy of 98.76\% on an independent test set, with F1 scores exceeding 95\% for each class. Applying the trained networks to the 27,335,836 KiDS DR5 targets with $r$-band magnitude brighter than 23 (i.e., $\rm MAG\_AUTO<23$), they obtained 6,549,300 stars, 2,804,654 quasars, and 17,981,882 galaxies. In this work, we select objects classified as galaxies with $r$-band apparent magnitudes brighter than 18.5, i.e., $\rm MAG\_AUTO<18.5$, yielding 310,583 galaxies. This magnitude threshold was deliberately chosen because brighter galaxies exhibit more pronounced morphological features and structural details, making them ideal candidates for our initial algorithm testing and validation. The obvious morphological features in brighter galaxies provide a more reliable foundation for developing our classification methodology before extending it to fainter objects with less distinct morphologies. 
    % Figure \ref{fig:4} presents the distributions of key physical parameters in our sample, including 
    % MAG\_AUTO measured by SExtractor \citep{Sextractor}, effective radius and Sérsic index $n$ determined by GaLNet \citep{GaLnet}, and stellar mass estimates from CIGALE \citep{mass}. The photometric redshift distribution, derived using GaZNet \citep{GaZnet}, is shown in Figure \ref{fig:z}.
    Figure \ref{fig:4} presents the distributions of key physical parameters, including 
    MAG\_AUTO measured by SExtractor \citep{Sextractor}, effective radius determined by GaLNet \citep{GaLnet}, photometric redshift distribution derived using the GaZNet \citep{GaZnet}, and stellar mass estimates from CIGALE \citep{mass}, shown for both our selected 
galaxy sample and the full KiDS dataset after applying a signal-to-noise 
ratio cut of S/N $>$ 35. 
    
%     Figure \ref{fig:4} presents the distributions of key physical parameters, including 
%     MAG\_AUTO measured by SExtractor \citep{Sextractor}, effective radius and 
%     Sérsic index $n$ determined by GaLNet \citep{GaLnet}, and stellar mass estimates from CIGALE \citep{mass}, shown for both our selected 
% galaxy sample and the full KiDS dataset after applying a signal-to-noise 
% ratio cut of S/N $>$ 35. 
% The photometric redshift distribution, derived using GaZNet \citep{GaZnet}, is shown in Figure \ref{fig:z}.  

\begin{table}[htbp]
    \centering
    \caption{Observational properties of the KiDS imaging data.}
    \label{tab:observation_propeties}
    \begin{tabular}{cccc}
    \hline
Band & Exp.time(s) & Mag. limit($5\sigma \ 2''$ AB) & Seeing(arcsec) \\
\hline
$u$ & 1000 & 24.8 & $0.9-1.1$ \\
$g$ & 900 & 25.4 & $0.7-0.9$ \\
$r$ & 1800 & 25.2 &  $< 0.7$\\
$i$ & 1200 & 24.2 &  $<1.1$\\
    \hline
    \end{tabular}
\end{table}

%\begin{table}[htbp]
    % \centering
%    \caption{Summary of the dataset used in this work.}
%    \label{table_data2}
%    \begin{tabular}{llll}
%    \hline
%    Property & Value \\
%\hline
%    Survey & KiDS DR5 \\
%Filters & $u$, $g$, $r$, $i$ \\
%Input & $g$, $r$, $i$ (RGB composite) \\
% Photometry & SExtractor MAG\_AUTO \\
%Magnitude cut & MAG\_AUTO $< 18.5$ \\
%Signal-to-noise cut & S/N $> 35$ \\
%Total sample size & 310{,}583$  \\
%Labeled subset & $5{,}000$  (five classes) \\
% Training set & $3{,}000$ galaxies \\
% Validation set & $1{,}000$ galaxies \\
% Test set & $1{,}000$ galaxies \\
%Unlabeled data &  $305{,}583$ \\
% Image size & $64 \times 64$ pixels \\
%    \hline
%    \end{tabular}
%\end{table}

\begin{figure}[htbp]
% \plotone{15.png}
\includegraphics[width=\columnwidth]{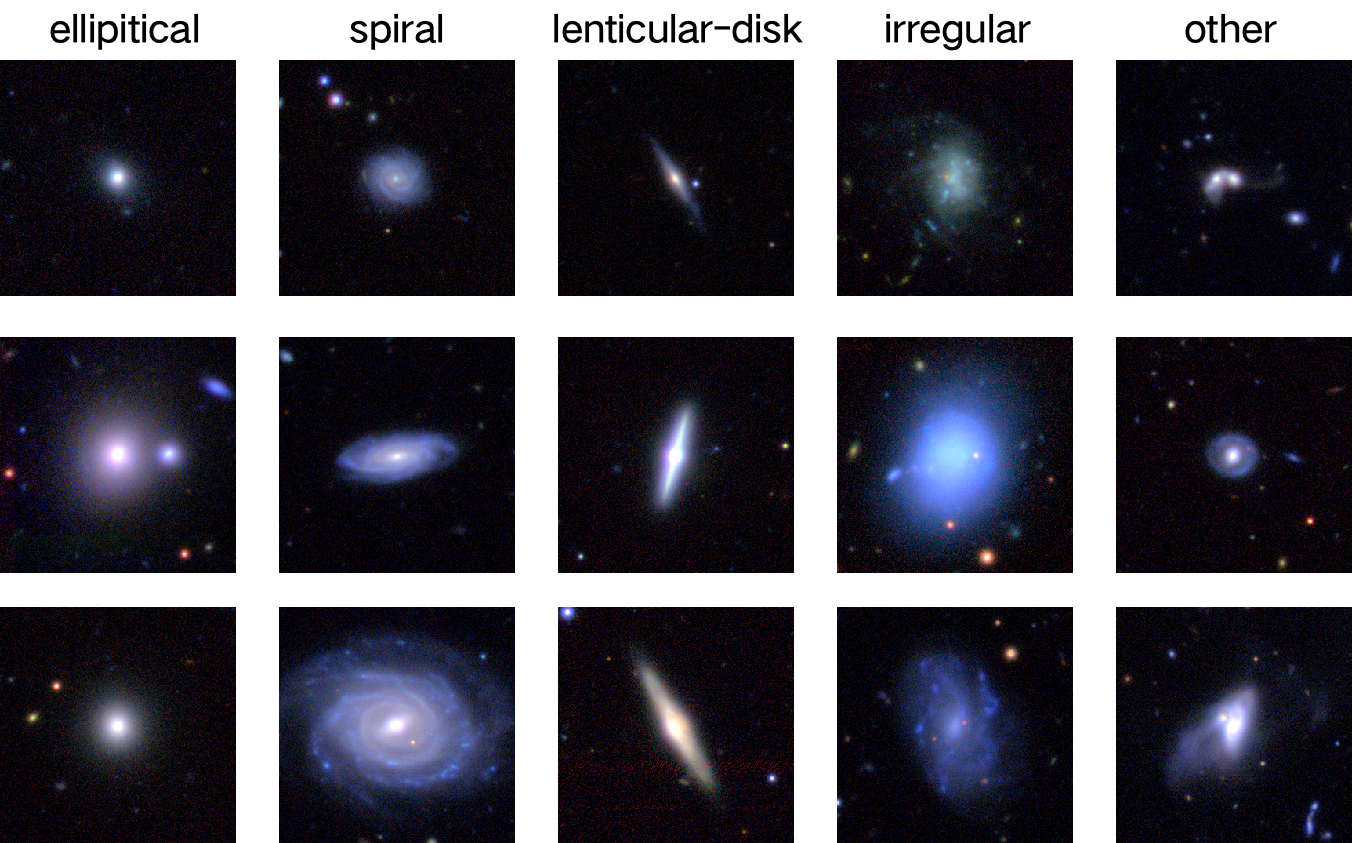}
    \caption{The $gri$ color-composite images of galaxy examples, where each column from left to right corresponds to a different morphological type: elliptical, spiral, lenticular-disk, irregular, and other galaxies. All images are displayed at a fixed size of $301 \times 301$ pixels. The pixel scale is $0.20''$/pixel.}
    \label{fig:15example}
\end{figure}

We classified the galaxies into five fundamental morphological categories: elliptical, spiral, lenticular-disk, irregular, and ``other'' galaxies. This classification scheme closely parallels the classical Hubble sequence \citep{hubble1926}, which has served as the cornerstone of galaxy morphological classification for nearly a century. Specifically, elliptical galaxies correspond to Hubble's E-type, spiral galaxies (both barred and unbarred) to Sa–Sd and SBa-SBd types, lenticular-disk galaxies to S0 types, and irregular galaxies to Irr types, while ``other'' includes galaxies that do not clearly fit into these categories. This approach provides a physically motivated framework that captures the most essential structural differences between galaxy types, facilitating subsequent comparative analysis of their properties and evolutionary pathways. Example images of each category are presented in Figure \ref{fig:15example}. 
    % Our complete dataset comprises approximately 310,000 galaxy images, with the majority reserved for encoder training. We randomly selected 5,000 images for manual classification, with an equal number of images (1,000 each) for the five morphological types. Of these, 60\% were allocated for classifier training, 20\% for validation, and 20\% for performance evaluation. 
    % These labeled images were removed from the complete dataset, leaving nearly 305,000 unlabeled images used for encoder training. This separation helps prevent bias caused by prior exposure to the labeled data.
Our complete dataset consists of 310,583 galaxy images.
% From this sample, we randomly selected 1,000 images for each of the five morphological types, totaling 5,000 labeled images. 
We randomly selected and visually inspected a subset of these galaxies, assigning them to five morphological classes. 
During the selection process, we required at least 12 arcseconds between the selected galaxies (to be labeled) and the remaining galaxies. This threshold was chosen to accommodate the 6 arcsecond random shift in our crop augmentation (see Section \ref{sec:structure}). Therefore, no matter how the crop is shifted, the target galaxy is always the closest one to the image center. The same isolation was also applied between the training and test splits of the labeled data. A galaxy that appears in the pretraining data may also appear in the labeled data, but only at the edge of the images. Since our model is trained and evaluated only on the central galaxy, these peripheral objects have no impact on our results.
%\textcolor{red}{And we ensured that no spatial overlap exists within 12 arcseconds between any labeled sample and the unlabeled set, nor between the training and test sets.} 
This resulted in a balanced labeled dataset of 5,000 galaxies (1,000 per class). We reserved 1,000 images (20\%) as a fixed test set, while the remaining 4,000 images were used for 5-fold cross-validation. In each fold, 3,200 images were used for training and 800 for validation.
The remaining 305,583 images, which are entirely unlabeled, were used exclusively for training the self-supervised encoder. This strict separation ensures that no images of the labeled dataset are used during the pre-training phase, thereby preventing any potential data leakage.

% \begin{figure}
% % \plotone{photoz.pdf}
% \includegraphics[width=\columnwidth]{./photoz_-0.1-5_widt0.1.pdf}
%     \caption{Photometric redshift distributions for the selected sample and the full KiDS dataset, with the vertical axis plotted on a logarithmic scale. The photometric redshifts are estimated using the GaZNet \citep{GaZnet} machine learning tool.}
%     \label{fig:z}
% \end{figure}

%
   % For a physical discussion of the stability criteria see Baker
   % (\cite{baker}) or Cox (\cite{cox}).

   % We observe that these criteria for dynamical, secular and
   % vibrational stability, respectively, can be factorized into
   % \begin{enumerate}
   %    \item a factor containing local timescales only,
   %    \item a factor containing only constitutive relations and
   %       their derivatives.
   % \end{enumerate}
   % The first factors, depending on only timescales, are positive
   % by definition. The signs of the left hand sides of the
   % inequalities~(\ref{ZSDynSta}), (\ref{ZSSecSta}) and (\ref{ZSVibSta})
   \section{Self-supervised Learning method} 
    \label{sec:method}
    In this section, we detail the self-supervised learning (SSL) method employed in our study. SSL leverages large amounts of unlabeled data to train models by designing pretext tasks that allow the model to learn meaningful feature representations without explicit supervision. Among various SSL approaches, contrastive learning has emerged as one of the most effective frameworks, demonstrating impressive performance across different domains. In our work, we employ a contrastive learning framework to pre-train the encoder, enabling the model to capture essential morphological features from galaxy images.

    \subsection{Structure}
    \label{sec:structure}
    The SSL process consists of two stages: a self-supervised pretraining stage and a supervised training stage. In the pretraining stage, the encoder learns to extract generalizable features from unlabeled data through a pretext task. In the supervised training stage, we freeze the encoder's parameters and add a classifier to map the learned features to specific galaxy morphology classes using a small set of labeled data. 
    Unlike traditional unsupervised learning relying on data intrinsic structure, SSL constructs pretext tasks with defined objectives (e.g., contrastive instance discrimination) for representation learning, while remaining label-free.
    % Unlike traditional unsupervised learning, which does not rely on any labels, SSL generates pseudo-labels through pretext tasks to guide the model's training process.
    
    In this study, we employ the SimCLR framework \citep{simclr2020}, a widely adopted contrastive learning approach. 
    This framework is built upon the ResNet50 architecture \citep{resnet_50}, which serves as the encoder (or feature extractor) in our work. We chose ResNet50 after carefully evaluating various backbone architectures, including ResNet18, ResNet34, and ResNet101 (results detailed in Section \ref{sec:network}). ResNet50 offers an optimal balance between model capacity and computational efficiency for our galaxy classification task. 
    While deeper networks like ResNet101 achieved marginally higher validation accuracy (0.2\%), they showed a 0.4\% degradation on the test set, and the substantial increase in training time and computational cost did not justify this modest improvement. 
% While deeper networks like ResNet101 achieved marginally higher accuracy (0.3\%), the substantial increase in training time and computational requirements did not justify this modest improvement. 
Conversely, while ResNet18 demonstrated competitive performance with significantly fewer parameters, ResNet50's additional capacity ensured better feature extraction for distinguishing subtle morphological differences in galaxy images, particularly for challenging classes like irregular galaxies.
    
    We remove the final classification layer (fully connected layer) of ResNet50 and attach a two-layer MLP as the projection head. The projection head maps the extracted features to a lower-dimensional space for contrastive learning, following the SimCLR framework. In the downstream task, we discard the projection head and replace it with a fully connected layer, which serves as the classifier for galaxy morphology classification. The SimCLR framework employs two distinct data augmentation techniques to create two different views (augmented versions) of the same image, which are treated as positive pairs in one batch. Correspondingly, the remaining samples in this batch are considered negative samples \citep{simclr2020}. 
    % Figure \ref{fig:augment} illustrates this process.

% \begin{figure*}[!htbp]
% % \plotone{K.pdf}
% \includegraphics[width=\textwidth]{BA2N.pdf}
%     \caption{After applying two different data augmentation processes, the original dataset with a batch size of $N$ is expanded to $2N$. Augmented images derived from the same original image form positive sample pairs, while those from different original images form negative sample pairs. These preprocessed images are $151 \times 151$ pixels.}
%     \label{fig:augment}
% \end{figure*}

    % \textcolor{red}{We did not apply additional preprocessing such as bad pixel removal or image-level normalization. However, we performed a standard center-cropping step to reduce the input size and computational cost while preserving the central galaxy before feeding the images into the model.}

In our work, we applied various augmentation techniques to galaxy images to ensure the augmented images retain the core morphological features while introducing enough variation to enhance the model's generalization ability during the pretraining phase. These augmentations include random translation, random 90° rotation, random horizontal flip, random vertical flip, color jitter, and random grayscaling. CenterCrop was applied afterward relative to the shifted image center to remove peripheral pixels (typically zero-padding) introduced by the translation and to standardize the input size, thereby preserving the translation-induced perturbations. 
% Figure \ref{fig:five} demonstrates these changes.

% \begin{figure*}
% % \plotone{augment.png} 
% \includegraphics[width=\textwidth]{Aug72.pdf}
%     \caption{The different changes after using distinct data augmentation. Note that center cropping is applied after augmentation to remove translated borders and standardize input dimensions, rather than serving as an augmentation itself. ``Composed'' means integrating all the previous steps. The uncropped augmented images are $151 \times 151$ pixels, while the center-cropped image and composed image are $64 \times 64$ pixels.} 
%     \label{fig:five} 
% \end{figure*}

\begin{itemize}
    \item random translation: translate the image randomly within a fixed range (20\%).
    \item random 90° rotation: rotate the image 90°, 180°, 270° or leave it unrotated (360° corresponds to no rotation), with a probability of 25\% each.
    \item random horizontal and vertical flip: flip the image horizontally or vertically with a probability of 50\% each.
    \item color jitter: randomly adjust the brightness, contrast, saturation, and hue of the image with a probability of 80\%.
    \item random gray-scaling: convert the image to grayscale with a probability of 20\%.
\end{itemize}
% The above changes do not include RandomResizedCrop in the original SimCLR paper, which generates interpolated pixels through resizing and cropping, thus creating new image content. In contrast, our augmentations do not involve geometric transformations such as random cropping with resizing or scaling, which, although not altering the intrinsic morphology of galaxies, are not physically meaningful in an astronomical context (e.g., scaling effectively corresponds to changing the redshift). They strictly preserve the original pixel grid, manipulating only existing pixel values without introducing artificial content.

\begin{figure*}[t]
% \plotone{Flow.png} 
\centering
\includegraphics[width=0.88\textwidth]{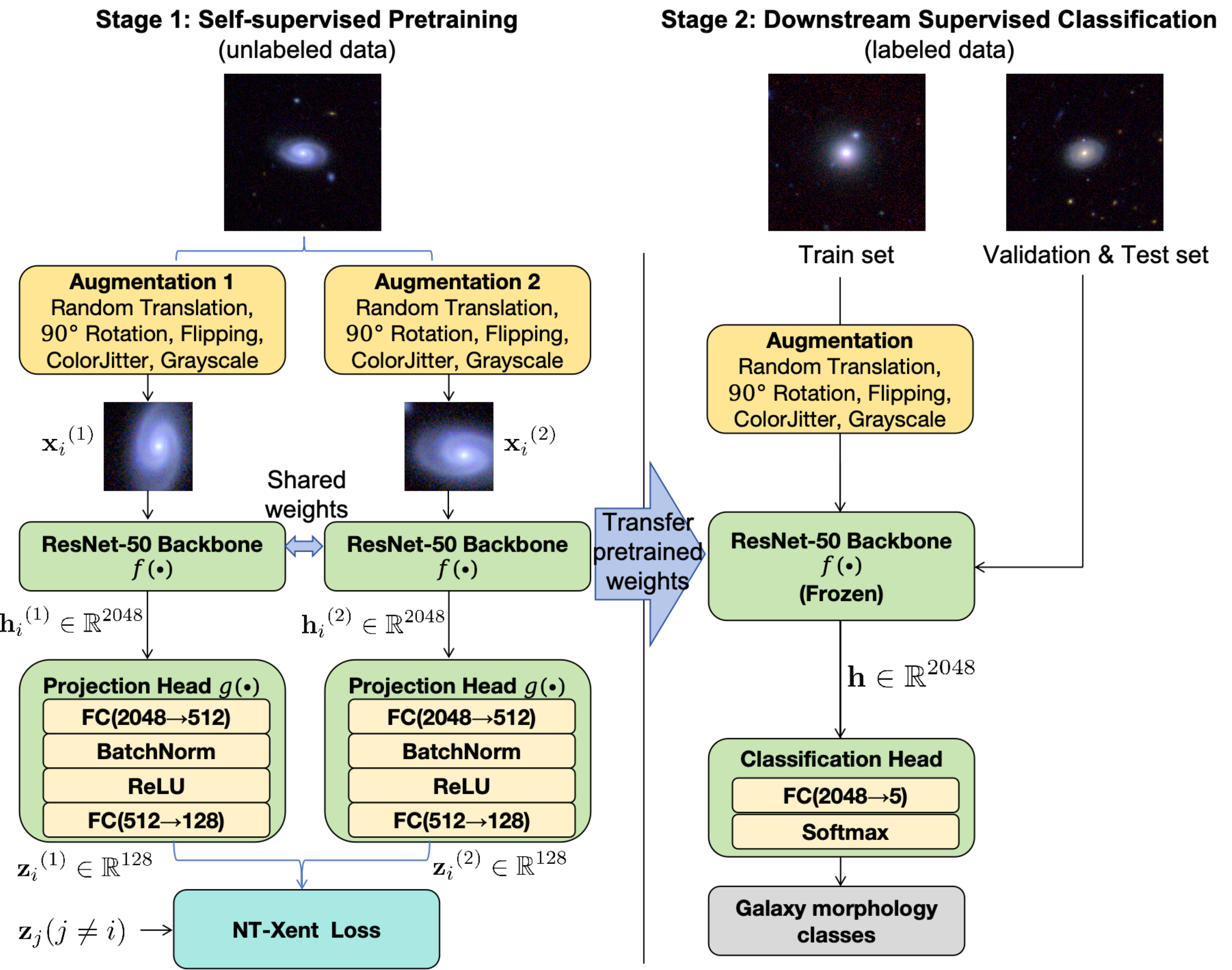}
    \caption{Flowchart of the overall process. During pretraining (left), augmented images are fed into the encoder and projection head to generate feature vectors for the NT‑Xent loss. In the downstream stage (right), the pretrained backbone is frozen and connected to a fully connected classifier for supervised galaxy morphology classification (shown in blue). Here, $z_j$ denotes the remaining feature vectors within the mini‑batch.} 
    \label{fig:flow} 
\end{figure*}

The whole structure is shown in Figure \ref{fig:flow}. Note that at this pretraining stage, we do not have any actual labels. Therefore, even though some images actually belong to the same class, if they are not derived from the same image, the model considers them as negative pairs. The model learns feature representations by maximizing the similarity between positive pairs while minimizing the similarity between negative pairs, using an NT-Xent contrastive loss function. The NT-Xent (Normalized Temperature-scaled Cross Entropy) loss function is formally defined as:
    \begin{equation} 
    \mathcal{L} = -\frac{1}{N}\sum_{i=1}^{N} \log \frac{\exp(\text{sim}(z_i, z_{i^+}) / \tau)}{\sum_{j=1}^{2N} \mathbf{1}_{[j \neq i]} \exp(\text{sim}(z_i, z_j) / \tau)},
    \label{eq:contrastive_loss}
    \end{equation}
where $z_i$ and $z_{i^+}$ are feature vectors from the same image under different augmentations, forming a positive pair. $\textit{sim}(z_i, z_{i^+})$ denotes cosine similarity of these two vectors, calculated as $\frac{z_i \cdot z_{i^+}}{||z_i|| \cdot ||z_{i^+}||}$. The parameter $\tau$ is the temperature scaling factor, where a lower $\tau$ increases contrast between positive and negative pairs \citep{tem}. The indicator function $1[j \neq i]$ ensures that the similarity computation excludes self-comparison.

\subsection{Linear evaluation protocol}
% It should be noted that we did not explicitly set a fixed random seed, so minor variations in results may occur due to stochastic initialization.
In our downstream task, we adopted a linear evaluation protocol, where the encoder is frozen and only a classifier is trained on top. The protocol has been widely used in self-supervised learning literature as a standard approach for evaluating learned representations, where the test accuracy serves as a direct proxy for the quality of the extracted features \citep{simclr2020}. In linear evaluation, data augmentation is significant for improving the model's generalization ability and accuracy. If we unfreeze some layers of the encoder or the entire encoder during downstream training, it becomes fine-tuning on a specific labeled dataset. Fine-tuning could lead to overfitting, especially with limited labeled data. By freezing the encoder, we reduce the risk of overfitting and maintain the model's ability to generalize to unseen data. Additionally, linear evaluation is more computationally efficient, as only the classifier is trained, reducing training time and resource consumption. Finally, linear evaluation demonstrates whether the self-supervised approach has effectively learned useful features from unlabeled data, without needing to retrain the entire model.
\subsection{Training process}
The preprocessing pipeline begins with original $301 \times 301$ pixel images, typically centered on the target galaxy. To reduce computational cost while preserving the main source, we initially center-crop these images to $151 \times 151$ pixels. We then apply data augmentation (including random translation, rotation, and flips) using similar strategies for both pretraining and downstream tasks. Finally, the images are center-cropped to $64 \times 64$ pixels to serve as network inputs. This input dimension was selected based on performance optimization; experiments with smaller ($44 \times 44$) or larger dimensions ($82 \times 82$, $101 \times 101$) yielded inferior results (see Section \ref{sec:cropsiz}). The overall workflow is illustrated in Figure \ref{fig:flow}.
During pretraining, the encoder processes the input pairs to generate 2048-dimensional feature vectors, which are mapped to a 128-dimensional space via a two-layer projection head to compute the contrastive loss. 
In the pretraining stage, the learning rate is set to $1 \times 10^{-3}$, weight decay is $1 \times 10^{-6}$, the batch size is 1024, and the number of epochs is 200.
For downstream classification, the model is trained for 200 epochs with a batch size of 64, a learning rate of $5 \times 10^{-5}$, and a weight decay of $1 \times 10^{-6}$. We employ the Adaptive Moment Estimation (Adam) optimizer \citep{adam}, which adaptively adjusts the learning rate based on first- and second-order gradient moments to ensure stable convergence and efficient optimization. The pretraining process requires approximately 5 hours on a single NVIDIA A100 GPU, representing a relatively moderate computational load compared to other deep learning applications.

Our approach demonstrates excellent scaling properties, making it suitable for even larger astronomical datasets. The self-supervised nature of the pretraining stage enables efficient utilization of the vast amounts of unlabeled data available in modern sky surveys. Additionally, once the encoder is trained, the downstream supervised training is highly efficient, requiring minimal labeled data and computational resources. This scalability makes our method particularly appealing for upcoming large-scale surveys like CSST and Euclid, which will produce unprecedented volumes of galaxy images.

\section{Performance} \label{sec:performance}
% loss and accuracy
In this section, we present a systematic assessment of our classification model's performance on the reserved test set. We begin by introducing the fundamental evaluation metrics essential for understanding classification effectiveness, followed by an analysis of the results obtained from our model.

Before the analysis, it is critical to acknowledge certain inherent limitations in our evaluation methodology. The ground truth labels utilized in this study are derived from visual classification by human experts, which introduces an unavoidable element of subjectivity into our assessment framework. Morphological classification, particularly for transitional or complex systems, remains challenging even for experienced astronomers. Certain galaxies occupy ambiguous regions in the morphological parameter space—such as S0/a galaxies at the lenticular-spiral boundary or peculiar systems with characteristics of multiple classes—making definitive categorization problematic even under ideal observing conditions. Furthermore, individual classifier biases, fatigue effects during extensive classification campaigns, and varying interpretations of the classification schema can contribute to inconsistencies in the training labels. These limitations inevitably influence our reported performance metrics, as the model may occasionally be penalized for ``misclassifications'' that actually represent a reasonable alternative interpretation of ambiguous morphologies. Despite these constraints, we adopt these expert-classified labels as our ground truth, as they represent the established standard in the field and provide the necessary foundation for supervised learning approaches. Future work might benefit from incorporating consensus classifications from multiple independent experts or probability distributions that better capture morphological uncertainty. Nevertheless, the high performance achieved by our model, particularly for well-defined morphological classes, indicates that these limitations do not substantially undermine the validity of our approach.

\subsection{Fundamental Classification Metrics}
The Receiver Operating Characteristic (ROC) curve represents a pivotal graphical tool for evaluating a classifier's discriminative ability across varying threshold settings. This curve plots the True Positive Rate (TPR) against the False Positive Rate (FPR) at different classification thresholds, providing comprehensive insight into the trade-off between sensitivity and specificity. The area under the ROC curve (AUC) quantifies overall classification performance, with values approaching 1.0 indicating superior discrimination capability.

For multi-class classification tasks such as galaxy morphology identification, several core metrics provide complementary perspectives on model effectiveness. In this context, we define positive samples as instances belonging to a specific class under consideration — for instance, galaxies exhibiting spiral features when evaluating the ``spiral" class performance. Correspondingly, negative samples lack the defining characteristics of the class being assessed.

The fundamental performance metrics are formally defined as follows:
\begin{equation}
\text{Accuracy} = \frac{\text{TP}+\text{TN}}{\text{TP}+\text{FP}+\text{TN}+\text{FN}}
\label{eq:acc}
\end{equation}

\begin{equation}
\text{Precision} = \frac{\text{TP}}{\text{TP}+\text{FP}}
\label{eq:precision}
\end{equation}

\begin{equation}
\text{Recall} = \frac{\text{TP}}{\text{TP}+\text{FN}}
\label{eq:recall}
\end{equation}

\begin{equation}
\text{F1-score} = 2 \times \frac{\text{Recall} \times \text{Precision}}{\text{Recall} + \text{Precision}}
\label{eq:f1}
\end{equation}

where TP (True Positives) represents correctly identified positive samples, TN (True Negatives) represents correctly classified negative samples, FP (False Positives) indicates negative samples incorrectly classified as positive, and FN (False Negatives) denotes positive samples erroneously classified as negative.

\subsection{Performance Analysis} \label{sec:4.2}
\begin{figure*}[htbp]
% \plotone{trainval.pdf}
\includegraphics[width=\textwidth]{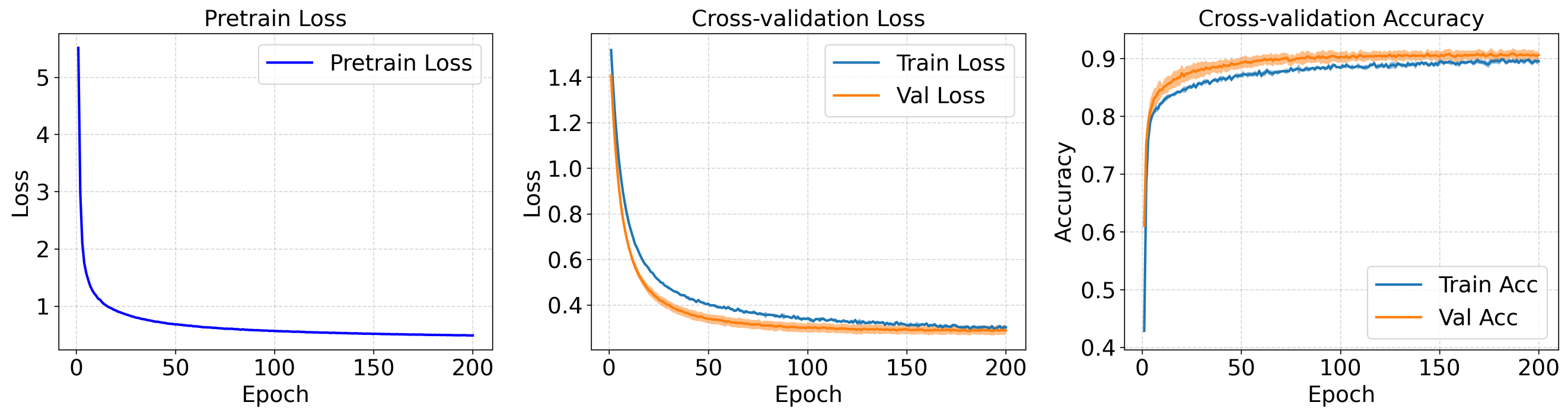}
\caption{The first panel is the pretraining loss curve during model optimization, showing the convergence trend across training epochs. The second panel shows the evolution of training loss (blue line) and validation loss (orange line) across epochs. The third panel displays the progression of training accuracy  (blue line) and validation accuracy (orange line).}
\label{fig:train-val-loss}
\end{figure*}

After completing the training process, we evaluated model performance by first analyzing the pretraining loss curve shown in the first panel of Figure \ref{fig:train-val-loss}. The loss exhibited a rapid decline during the initial 25 epochs, followed by a more gradual but consistent decrease in subsequent epochs. After 200 epochs, the loss curve plateaued, indicating convergence and suggesting that the network had reached optimal training. This convergence pattern confirms that the pretraining phase was sufficiently long to allow the model to learn robust feature representations before proceeding to downstream tasks. The second and third panels of Figure \ref{fig:train-val-loss} show the loss and accuracy metrics observed during the downstream classification training process. The training and validation loss curves demonstrate nearly identical declining trajectories and exhibit substantial overlap, indicating the absence of overfitting phenomena. After 200 epochs, the loss function converges to an optimal state. The corresponding accuracy metrics, as shown in the third panel of Figure \ref{fig:train-val-loss}, achieve approximately 90\% for both training and validation datasets, further confirming the model's robust capabilities.

% \begin{figure}
% % \plotone{preloss.pdf}
% \includegraphics[width=\columnwidth]{preloss.pdf}
%     \caption{The pretrain loss curve during model optimization, showing the convergence trend across training epochs.}
%     \label{fig:preloss}
% \end{figure}

% \begin{figure}[ht!]
% \centering
% % \plotone{rC2.pdf}
% \vspace{0.5cm}
% \includegraphics[width=\columnwidth]{./rC200-200_5e5.pdf}
%     \caption{The ROC curves for five galaxy types, with the false positive rate (FPR) and true positive rate (TPR) plotted on logarithmic scales. The curves represent
% five morphological classes: Elliptical (blue), Spiral (orange), Irregular (green), Lenticular-Disk
% (red) and Other (purple).}
%     \label{fig:roc curve}
%     % \vspace{0.3cm}
% \end{figure}

\begin{figure*}[htbp]
\centering
% \plotone{con.pdf}
\includegraphics[width=0.9\textwidth]{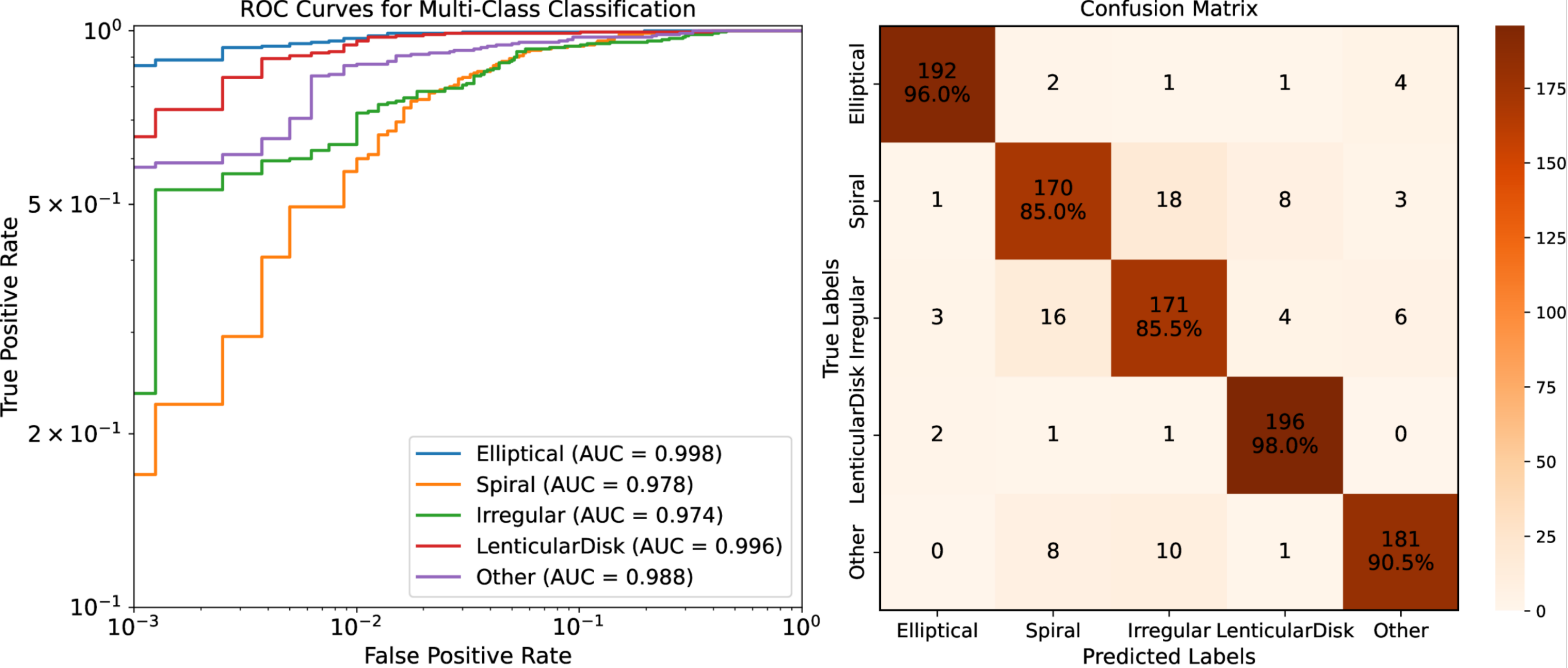}

\caption{The left panel shows the ROC curves for five galaxy types, with the false positive rate (FPR) and true positive rate (TPR) plotted on logarithmic scales. The curves represent
five morphological classes: Elliptical (blue), Spiral (orange), Irregular (green), Lenticular-Disk
(red) and Other (purple). The right panel is the confusion matrix for galaxy morphology classification across five types: elliptical, spiral, irregular, lenticular-disk, and other. Rows show true labels, columns show predicted labels. Diagonal elements (darkest colored) indicate correct predictions, with color intensity scaling proportionally to classification recall for each class.}
\label{fig: confusion matrix}
% \vspace{0.2cm}
\end{figure*}

Following the supervised training phase, we evaluated our model using the reserved test set.  For visualization and subsequent analysis, we present the results obtained using the best-performing model. The left panel of Figure \ref{fig: confusion matrix} illustrates the ROC curves for the five galaxy morphological types. The logarithmic scale on both axes enhances visibility of the high-performance regions. All classes demonstrate exceptional discriminative capability, with AUC values of 0.998 for elliptical and 0.996 for lenticular-disk galaxies, 0.988 for the ``other'' category, 0.978 for spiral galaxies, and 0.974 for irregular galaxies. These AUC values demonstrate the model’s strong discriminatory performance on the KiDS DR5 test set.
% These near-perfect AUC values confirm the model's robust ability to differentiate between morphological types across different classification thresholds. 
The right panel of Figure \ref{fig: confusion matrix} presents the confusion matrix obtained from the test set, providing a detailed assessment of classification performance for each galaxy type. The predicted class for each galaxy is assigned to the label with the highest probability. The matrix reveals that elliptical and lenticular-disk galaxies achieve the highest classification recall (96\% and 98\%, respectively), while spiral and irregular galaxies show slightly lower but still impressive recall rates (85\% and 85.5\%, respectively). The ``other'' category achieves 90.5\% recall despite its inherent heterogeneity. The diagonal predominance in the confusion matrix further confirms the model's strong classification capabilities across all morphological types.

\begin{figure*}[!htb]
     \centering
     \includegraphics[width=0.9\textwidth]{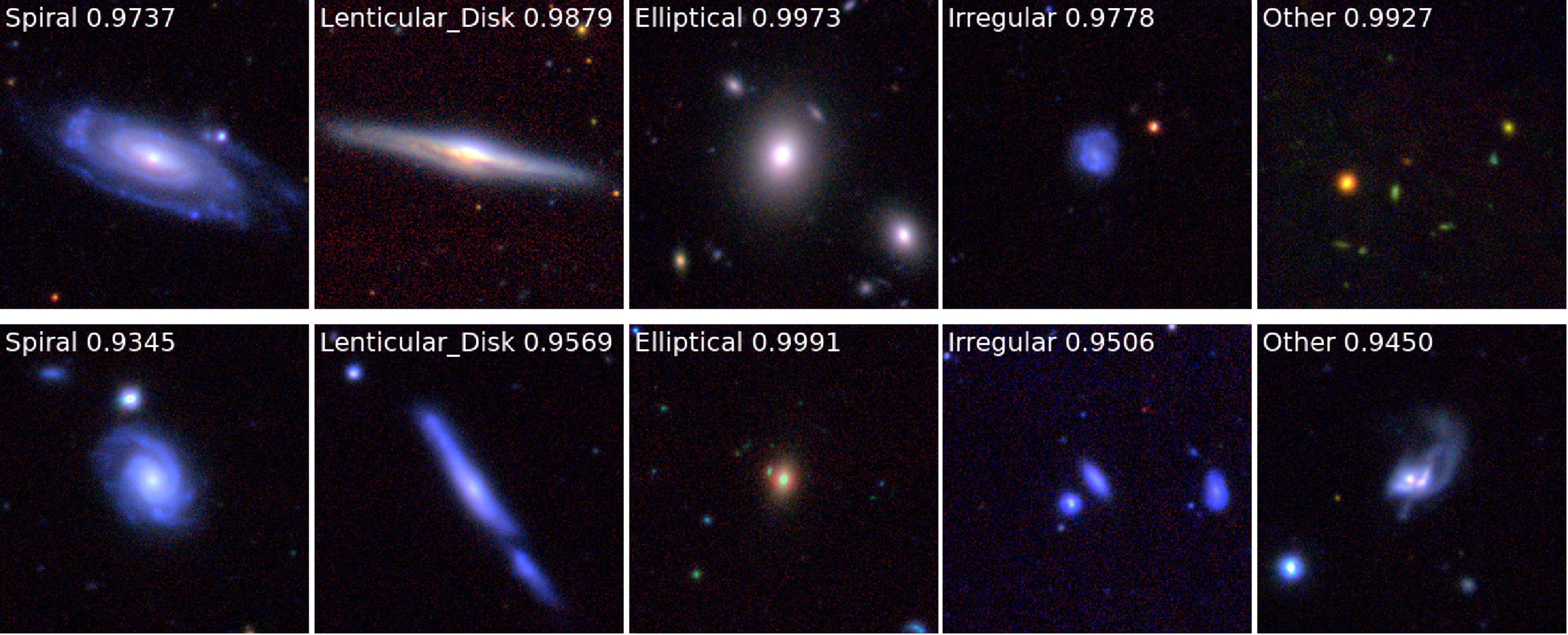}
    \caption{Some examples of successful classifications. The top left corner of each image indicates the predicted category and the probability by the model. Image size: $301 \times 301$ pixels.} 
    \label{fig:right} 
\end{figure*}

\begin{figure*}
    \centering
     \includegraphics[width=0.9\textwidth]{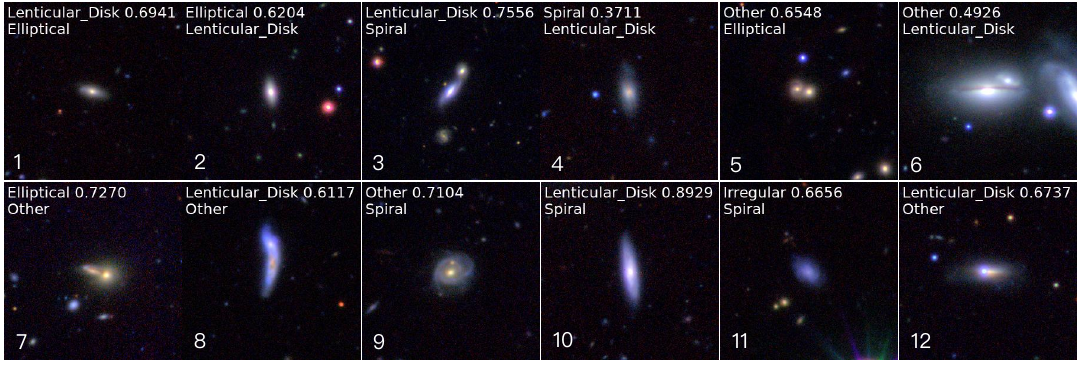}
    \caption{Examples of failed classifications. 
    The top left corner of each image indicates the predicted category and the probability by the model, while the true label is displayed below for reference. 
    Image size: $301 \times 301$ pixels.}
    \label{fig:wrong} 
\end{figure*}

Figure~\ref{fig:right} shows example cases of successful galaxy classification. For each image, the top-left corner lists the predicted category and its probability. In general, successful classifications correspond to clear morphological signatures: from left to right, the panels show well-defined spiral arms, lenticular features, smooth elliptical light profiles, irregular substructures, and, in the last column, ambiguous targets that are hard to classify. 
%Figure~\ref{fig:wrong} presents cases where the model predictions disagree with the true morphological labels. These misclassifications typically occur for galaxies with ambiguous morphologies or with features that resemble multiple classes.

Elliptical and lenticular-disk galaxies generally exhibit more symmetric and regular morphologies, leading to higher classification recalls (96\% and 98\%, respectively). Their consistent visual patterns facilitate more reliable feature extraction and classification. However, distinguishing between lenticular-disk and elliptical galaxies can be challenging due to projection effects (see Figure \ref{fig:wrong}, id=1,2). Although lenticular-disk galaxies are intrinsically disk systems, they may lack obvious disk features when viewed face-on, leading to misclassification as ellipticals. Conversely, in edge-on projection, they can resemble elongated ellipticals. Furthermore, some elliptical galaxies with low ellipticity may appear similar to edge-on lenticular-disk galaxies, further complicating their differentiation. Another particularly challenging failure mode involves edge-on spiral galaxies, which present fundamentally different morphological signatures from their face-on counterparts (Figure \ref{fig:wrong}, id=3,4). When a spiral galaxy's disk is oriented nearly perpendicular to our line of sight (inclination$>$80$^\circ$), its spiral arms—typically the most diagnostic feature—become compressed into a thin, elongated structure. In such orientations, the galaxy's light profile is dominated by the disk and bulge, closely mimicking lenticular-disk or even elliptical morphologies.

Interestingly, the ``other" category achieves a relatively high recall (90.5\%), despite encompassing diverse objects such as galaxy clusters, ring galaxies, and mergers.
%(Figure~\ref{fig:find}). 
This strong performance likely reflects the fact that these objects exhibit clear visual differences from the main morphological classes (elliptical, spiral, lenticular-disk, and irregular), allowing the model to distinguish them largely by exclusion. However, within this category, merging and interacting galaxies present a particular challenge. These systems inherently defy the Hubble sequence paradigm, transitioning through morphologically ambiguous intermediate states that combine features of multiple types. Our visual inspection of the test set reveals that 28.5\% of the ``other" category classifications (57 systems) are ongoing or recent mergers, identified by asymmetric light distributions, tidal tails, and multiple nuclei.
Notably, the model achieves reasonable classification of merger systems despite their inherent morphological complexity, with 48 classified as ``other'' (84.2\%).
The classification performance on these merger systems is lower than that for classical galaxy types. This difficulty often arises in systems containing multiple galaxies: the model may classify a single galaxy within the system (assigning it a normal type such as elliptical or spiral) or the full system as a whole (assigning it as a merger or interaction) (Figure \ref{fig:wrong}, id=5, 6, 7, 8). This naturally creates bidirectional confusion—some mergers are classified as normal galaxies, and some normal-looking galaxies are classified as mergers. Such behavior is understandable: the model has no built-in notion of whether a system should be treated as a whole, so its label depends on which visual pattern dominates the input. More fundamentally, the morphological diversity of mergers—ranging from early-stage interactions with minimal disturbance to advanced mergers with completely disrupted structures—makes consistent classification challenging. This difficulty is compounded by the limited number of training examples (243 out of 1,000 ``other" samples), which cannot adequately represent the full range of merger stages.

Some images show potential label errors—cases where the model's predictions are more consistent with the observed morphology than the original labels. Representative examples are presented in the last 4 samples (id=9, 10, 11, 12) of Figure~\ref{fig:wrong}. These discrepancies likely arise from visual ambiguity or overlooked structural features during the initial labeling process. For example, some galaxies show subtle merger-related features, while others display faint and ambiguous features (e.g., faint spiral arms or bars), all of which can complicate visual inspection. More generally, weak or diffuse light distributions and indistinct morphological boundaries can also introduce annotation inconsistencies, contributing to label noise that degrades classifier performance.

\section{Analysis of Model Configuration} \label{sec: ablation}

In this section, we first evaluate the effectiveness of our self-supervised pretraining by comparing it with training from scratch. Then we investigate the different settings in the training process, including the amount of labeled data, crop sizes and model backbones.

% \subsection{Impact of Labeled Data Size}
% \textcolor{blue}{We evaluate the effect of labeled data set size on classification performance. Using ResNet50 as the backbone, we vary the number of labeled images from 500 to 4,000 (500, 1,000, 2,000, 3,000, 4,000). For each data size, we evaluated with 5-fold cross-validation on a fixed test set of 1,000 images.
% As shown in Figure \ref{fig:5fold}, both validation and test accuracy increase with larger training data. Test accuracy saturates after 3,000 samples, while validation accuracy continues to increase slowly.
% }

% \begin{figure}
%     \centering
%     \includegraphics[width=\columnwidth]{./5fold4.pdf}
%     \caption{Impact of labeled data size on model performance. Validation and test accuracy improve as the number of labeled images increases from 500 to 4,000. Test accuracy stabilizes beyond 3,000 samples, indicating diminishing returns from additional labeled data.}
%     \label{fig:5fold}
% \end{figure}

\subsection{The Impact of Self-Supervised Pretraining}

To evaluate the contribution of self-supervised learning to our performance, we conducted a controlled comparison between two models with identical architectures that differed only in the initialization of the encoder weights. The experimental group used encoder weights obtained from the self-supervised pretraining stage, while the control group used randomly initialized weights. To ensure a direct and fair comparison, all other aspects of the experiment—including the network architecture, optimizer settings, learning rate schedule, and data augmentation—were held constant for both groups. Both models were then trained on the same limited dataset of 4,000 labeled images and evaluated on a fixed test set of 1,000 images.

We found that the model trained from scratch performed poorly, highlighting the fundamental difficulty of training deep neural networks when labeled data are scarce. The training process itself was highly unstable. Rather than decreasing steadily, the training loss exhibited large and erratic fluctuations, and the validation accuracy stagnated without clear improvement, eventually plateauing near 40\%. This poor outcome reflects the difficulty of optimizing a ResNet encoder with millions of parameters when only 4,000 labeled examples are available—the gradient signal is insufficient to guide the model toward a good solution. In sharp contrast, the model initialized with pretrained weights showed far better training behavior and final performance. From the start of the supervised training, this model was remarkably stable. The training loss decreased smoothly and consistently, and the validation accuracy improved steadily and reliably, indicating that the model was effectively converging to a good solution. This difference arises because the self-supervised pretraining phase provides an effective starting point. By learning from a large collection of unlabeled data, the encoder has already captured a rich and broadly applicable set of visual features—from simple patterns to the more complex structures of galaxies.

This comparison demonstrates that self-supervised pretraining is critical to our approach. By leveraging large amounts of unlabeled survey data, it enables stable training and strong performance where training from scratch fails. For astronomy, where large labeled datasets are expensive and time-consuming to produce, self-supervised learning bridges the gap between the data requirements of deep neural networks and the practical constraints on labeled data availability.
%This initial learning process guides the encoder to a very promising region within the vast parameter space. As a result, the main training task changes from a very difficult optimization problem to a much simpler one:
%\textcolor{blue}{either keeping the pretrained encoder fixed and training a simple classifier on top of its features, or fine-tuning the entire network from this strong initialization—both of which become feasible and effective with the rich representations learned during self-supervised pretraining.}
% keeping the pretrained encoder fixed and training a simple classifier to link these meaningful features to the final categories.

\subsection{Impact of Labeled Data Size}
\begin{figure*}
    \centering
    \includegraphics[width=\textwidth]{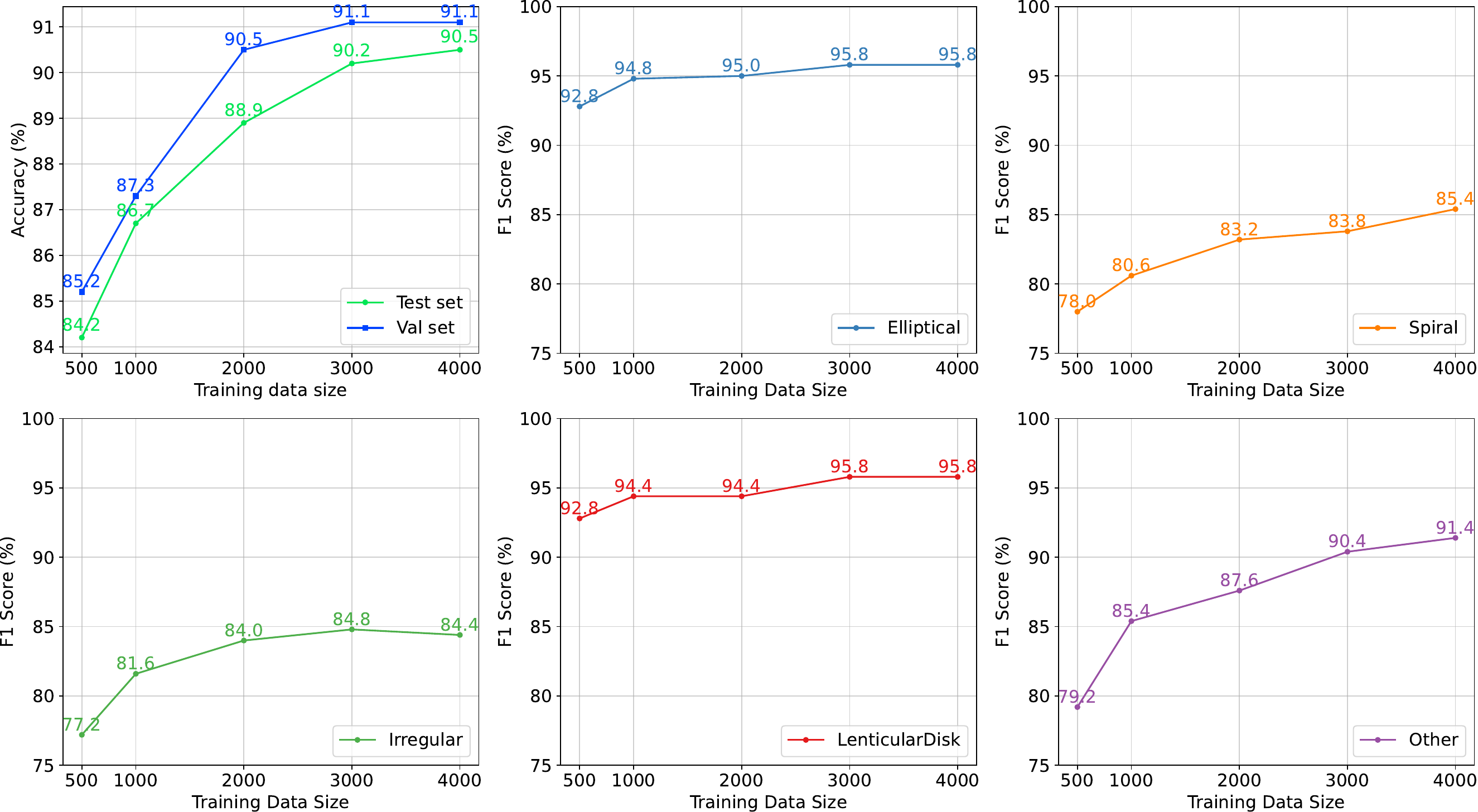}
    \caption{Impact of labeled data size on model performance. Top-left: validation and test accuracy improve as the number of labeled images increases from 500 to 4,000. Test accuracy stabilizes beyond 3,000 samples. Other subplots: F1 scores for the five galaxy types. Elliptical and lenticular-disk galaxies reach near-saturated performance with relatively small datasets, whereas spiral, irregular, and ``other" types benefit more from larger labeled samples.}
    \label{fig:5fold}
\end{figure*}

We evaluated the effect of labeled dataset size on classification performance. Using ResNet50 as the backbone, we varied the number of labeled images from 500 to 4,000. For each data size, we performed 5-fold cross-validation with a fixed test set of 1,000 images. As shown in the top-left panel of Figure~\ref{fig:5fold}, both validation and test accuracy increase with larger training data. The improvement is most pronounced in the low-data regime: increasing from 500 to 2,000 images yields a gain of 4.7 percentage points. Beyond 2,000 images, the gains diminish sharply, with only a 1.6 percentage point improvement from 2,000 to 4,000 images. Validation accuracy saturates after 3,000 samples, while test accuracy continues to increase slowly. This trend suggests that the representations learned during self-supervised pretraining already contain substantial morphological information from the large unlabeled dataset. As the number of labeled images increases, the classifier can better adapt these representations to the classification task. Once sufficient labeled data are available, the marginal benefit of additional samples decreases, producing the observed performance saturation.

In the other subplots of Figure~\ref{fig:5fold}, the F1 scores for the five galaxy types also improve consistently. Notably, for elliptical and lenticular-disk galaxies, the scores are nearly saturated at larger data sizes, reflecting their symmetric and regular morphologies. In contrast, for spiral, irregular, and ``other'' types, F1 scores improve significantly as data size increases. Unlike elliptical and lenticular-disk galaxies, these classes exhibit more diverse morphological features. Larger labeled datasets provide more representative instances of such complex morphologies, enabling the model to learn more discriminative features for challenging cases. Among all classes, irregular galaxies exhibit the lowest F1 values across all data scales, which directly reflects their extreme morphological variability. The ``other'' category, which aggregates various transitional and merging galaxy subtypes, shows the largest performance boost: its F1 score increases from 79.2\% (500 samples) to 91.4\% (4,000 samples). With expanded training data, the model encounters more heterogeneous transitional galaxy features and learns to distinguish these ambiguous subtypes more reliably, yielding steadily improving F1 scores.
These findings highlight the data efficiency of the proposed framework. For large astronomical surveys, where unlabeled observations are abundant but reliable visual classifications are expensive to obtain, self-supervised learning provides an effective way to reduce the need for large labeled datasets while maintaining strong classification performance.

% \subsection{Input Optimization and Network Model Configuration} \label{sec:network}
% In this section, we detail the experiments conducted to determine the optimal configuration for our model. We first investigate the effect of input image crop size to establish the best data preprocessing strategy. Then, using this optimal crop size, we evaluate a series of ResNet architectures to select the most suitable backbone for our galaxy classification task.
\subsection{Input Optimization}\label{sec:cropsiz}
Although our images did not undergo extensive preprocessing, the choice of central crop size constitutes an important preprocessing step that directly affects the amount of morphological information available to the model.
The input crop size is a critical hyperparameter that dictates the information available to the model. Using ResNet50 as the backbone, we experimented with several central crop sizes: 44, 64, 82, and 101 pixels. Our findings, summarized in Table \ref{tab:crops}, show that the crop size has a significant impact on performance. A large crop of 101 pixels proved to be counterproductive, yielding a relatively low accuracy of approximately 83\%. We attribute this to the inclusion of excessive background noise, which can interfere with the model's ability to learn discriminative features from the central galaxy. In contrast, smaller crop sizes performed much better, with final test accuracies of 89\% (44 pixels), 88\% (82 pixels), and a peak performance of 91\% (64 pixels). This suggests that the core region of the images contains the most vital morphological information, which aligns with the fact that most galaxies in our dataset are relatively small and centered. Based on these results, we selected a $64 \times 64$ pixel central crop as the standard for our subsequent experiments. However, it is important to acknowledge the limitations of this fixed-size cropping strategy. We observed that for unusually large galaxies, a $64 \times 64$ pixel crop can exclude significant portions of the galaxy's structure, leading to misclassification. An alternative is to resize the entire image to $64 \times 64$ pixels, preserving more information for these large objects and leading to correct classification. While central cropping provides the best overall performance across the dataset, future work could explore adaptive strategies that combine cropping and resizing to improve robustness for galaxies with large size variations.

%\begin{figure}[h]
%     \includegraphics[width=\columnwidth]{./difcrop.pdf}
%    \caption{The image above uses center cropping with a size of 64, while the one below is scaled to 64. For clarity, the processed images have been enlarged. As can be seen, center cropping loses many features, leading to incorrect classifications by the model. Original image and two classified images: $301 \times 301$ pixels. Two processed images: $64 \times 64$ pixels.} 
%    \label{fig:different} 
%\end{figure}

\subsection{Network Model Configuration}\label{sec:network}
Using the optimized 64-pixel crop size, we evaluated the performance of four residual network architectures: ResNet18, ResNet34, ResNet50, and ResNet101. Each encoder was initialized with pretrained weights from a dedicated self-supervised pretraining stage using the same architecture, ensuring a fair comparison. Using 4,000 labeled images for training and a fixed test set of 1,000 images, we performed 5-fold cross-validation for each model. Table~\ref{tab:resnet_performance} shows the results. ResNet101 achieved the highest validation accuracy (91.3\%), with ResNet50 performing very similarly (91.1\%). However, on the test set, ResNet50 outperformed ResNet101 (90.5\% vs. 90.1\%). 
For both architectures, the validation accuracy slightly exceeds the test accuracy. Given that these gaps (0.6\% and 1.2\%, respectively) are at the level of the standard deviation of the cross-validation folds, we attribute this minor discrepancy to random sampling fluctuations in data partitioning rather than a systematic domain shift.

We noticed that, for both of the models, the validation accuracy is higher than the test accuracy. Given that this gap is smaller than the corresponding standard deviation of the cross-validation folds, we suggest that this minor difference simply reflects random data-splitting variation rather than a systematic domain shift.

Notably, the substantially lighter ResNet18 also performed well, achieving a test accuracy of 89.5\%. This suggests that even a shallower network has sufficient capacity to extract the key features for this classification task, offering a significant advantage in terms of computational efficiency and training time. The performance gains from using deeper models like ResNet50 and ResNet101 were relatively small compared to their increased computational cost. Considering this trade-off between performance and efficiency, we selected ResNet50 as the primary backbone for our main experiments. It delivers near-peak accuracy while being more computationally practical than deeper models, making it a balanced choice for this work. Further analysis of detailed metrics (precision, recall, F1-score), provided in Table \ref{tab:resnet_performance}, shows that these scores are quite similar across all tested models. We note, however, that scores for spiral and irregular galaxies are consistently lower than those for elliptical and lenticular-disk galaxies. This is likely because the latter two types have more regular and easily identifiable features. The stability of these metrics across different network depths further supports the conclusion that shallower ResNets are already powerful enough to capture the essential characteristics of these galaxy types.

\begin{table}[h]
    %\centering
    \caption{Classification performance across different crop sizes.}
    \begin{tabular}{l|c c c c}  
        \hline
        \hline
        size & galaxy type & precision & recall & F1-score \\  
        \hline
         {44} 
        & Elliptical  &  0.97  &  0.96 &  0.97  \\  
        & Spiral &  0.85  &  0.82  &  0.84  \\
        & Irregular  &  0.84  &  0.83  &  0.84  \\
        & Lenticular-Disk    &  0.93  &  0.97  &  0.95  \\
        & Other    &  0.89  &  0.90  &  0.90  \\ 
        \hline
        {64}  
        & Elliptical  &  0.97  &  0.96  &  0.96  \\  
        & Spiral &  0.86  &  0.85  &  0.86  \\
        & Irregular  &  0.85  &  0.86  &  0.86  \\
        & Lenticular-Disk    &  0.93  &  0.98  &  0.95  \\
        & Other    &  0.93   &  0.90  &  0.92  \\
        \hline
        {82} 
        & Elliptical  &  0.96  &  0.95 &  0.96  \\  
        & Spiral &  0.84  &  0.82  &  0.83  \\
        & Irregular  &  0.78  &  0.78  &  0.78  \\
        & Lenticular-Disk    &  0.92  &  0.96  &  0.94  \\
        & Other    &  0.88  &  0.88  &  0.88  \\ 
        \hline
        {101} 
        & Elliptical  &  0.94  &  0.94 &  0.94  \\  
        & Spiral &  0.77  &  0.79  &  0.78  \\
        & Irregular  &  0.70  &  0.66  &  0.68  \\
        & Lenticular-Disk    &  0.90  &  0.95  &  0.92  \\
        & Other    &  0.86  &  0.84  &  0.85  \\ 
        \hline
        \hline
    \end{tabular}
    \textit{Note.} Column 1 represents the crop size of pixels. Columns~2 through~5 present galaxy-type labels together with the corresponding precision, recall, and F1-score for each class under different crop sizes.
    \label{tab:crops}
\end{table}

\renewcommand{\dblfloatpagefraction}{.9}
% \begin{table*}[t]
%     \caption{Validation and test accuracy of different ResNet models.}
%     \centering
%     \small
%     % \setlength{\tabcolsep}{3pt}
%     \begin{tabular}{l|c c c c}
%         \hline
%         Data & ResNet18 & ResNet34 & ResNet50 & ResNet101\\  
%         \hline
%         Val  &  $89.9\%\pm0.6\%$  &  $89.6\%\pm1.0\%$  &  $91.1\%\pm0.7\%$ &  $91.3\%\pm0.8\%$\\  
%         Test &  $89.5\%\pm0.5\%$  &  $89.1\%\pm0.4\%$  &  $90.5\%\pm0.2\%$ &  $90.1\%\pm0.3\%$\\
%         \hline
%     \end{tabular}
%     % \caption{Validation and test accuracy of different ResNet models. `Val' represents the validation. While deeper models generally outperform shallower ones, ResNet34 shows degraded test accuracy despite comparable validation performance ($88.5\%\pm0.3\%$ vs. ResNet18's $88.6\%\pm0.2\%$), suggesting potential hyperparameter sensitivity. The accuracy is average value of the last 10 epochs with the standard deviation.}
    
%     % \vspace{2mm}
%     {\raggedright
% \textit{Note.} 
% \textcolor{blue}{
% `Val' represents the validation. While deeper models generally outperform shallower ones, ResNet34 shows degraded test accuracy despite comparable validation performance ($89.9\%\pm0.6\%$ vs. ResNet18's $89.6\%\pm1.0\%$). A similar trend is observed for ResNet101, which achieves comparable validation accuracy to ResNet50 but slightly lower test accuracy ($90.1\%\pm0.3\%$ vs. $90.5\%\pm0.2\%$).
% }\par
% }
%     \label{tab:per}
% \end{table*}

\begin{table*}[t]
    \centering
    \caption{
    Performance comparison for different ResNet depths.
    % Classification performance (precision, recall, and F1-score) of different ResNet architectures.
    }
    \label{tab:resnet_performance}

    \renewcommand{\arraystretch}{1.2}
    \begin{tabular}{l| c| c |l c c c}
        \hline
        \hline
        model & validation & test & galaxy type & precision & recall & F1-score \\
        \hline

        \multirow{5}{*}{ResNet18} 
        & \multirow{5}{*}{$89.9\%\pm0.6\%$} 
        & \multirow{5}{*}{$89.5\%\pm0.5\%$}
        & Elliptical        & 0.96 & 0.96 & 0.96 \\
        & & & Spiral             & 0.86 & 0.81 & 0.83 \\
        & & & Irregular          & 0.83 & 0.88 & 0.85 \\
        & & & Lenticular-Disk    & 0.92 & 0.97 & 0.95 \\
        & & & Other              & 0.93 & 0.90 & 0.91 \\
        \hline

        \multirow{5}{*}{ResNet34} 
        & \multirow{5}{*}{$89.6\%\pm1.0\%$} 
        & \multirow{5}{*}{$89.1\%\pm0.4\%$}
        & Elliptical        & 0.95 & 0.95 & 0.96 \\
        & & & Spiral             & 0.82 & 0.84 & 0.83 \\
        & & & Irregular          & 0.86 & 0.80 & 0.83 \\
        & & & Lenticular-Disk    & 0.92 & 0.98 & 0.95 \\
        & & & Other              & 0.91 & 0.89 & 0.90 \\
        \hline

        \multirow{5}{*}{ResNet50} 
        & \multirow{5}{*}{$91.1\%\pm0.7\%$} 
        & \multirow{5}{*}{$90.5\%\pm0.2\%$}
        & Elliptical        & 0.97 & 0.96 & 0.96 \\
        & & & Spiral             & 0.86 & 0.85 & 0.86 \\
        & & & Irregular          & 0.85 & 0.86 & 0.86 \\
        & & & Lenticular-Disk    & 0.93 & 0.98 & 0.95 \\
        & & & Other              & 0.93 & 0.90 & 0.92 \\
        \hline

        \multirow{5}{*}{ResNet101} 
        & \multirow{5}{*}{$91.3\%\pm0.8\%$} 
        & \multirow{5}{*}{$90.1\%\pm0.3\%$}
        & Elliptical        & 0.96 & 0.97 & 0.97 \\
        & & & Spiral             & 0.85 & 0.85 & 0.85 \\
        & & & Irregular          & 0.87 & 0.83 & 0.85 \\
        & & & Lenticular-Disk    & 0.93 & 0.98 & 0.95 \\
        & & & Other              & 0.92 & 0.91 & 0.91 \\
        \hline
        \hline
    \end{tabular}
    
{\raggedright
\textit{Note.} 
Column 1 lists the ResNet variants, where the suffix indicates the number of convolutional layers in each architecture. Columns~2 and~3 report the overall accuracy (mean $\pm$ standard deviation) on the validation and test sets, respectively. Columns~4 through~7 present galaxy-type labels together with the corresponding precision, recall, and F1-score for each class under every model. %\textcolor{red}{The validation accuracies are slightly higher than the test accuracies; this small difference is within the expected variation arising from the random sampling of the validation and test sets.}
\par
}
\end{table*}

\section{Applying to KIDS DR5} \label{sec:apply}

Following the supervised classifier training phase,
% fine-tuning phase,
we applied the optimized classifier to categorize the entire dataset that was initially utilized during the self-supervised pre-training stage. This classification yielded 115,136 elliptical galaxies (37.7\%), 86,895 spiral galaxies (28.4\%), 42,304 irregular galaxies (13.8\%), 37,980 lenticular-disk galaxies (12.4\%), and 23,268 galaxies (7.7\%) assigned to the ``other" category. This categorical distribution aligns well with the established understanding of galaxy demographics in the local universe. Ellipticals and spiral galaxies constitute the majority of the population, followed by irregular and lenticular systems, while specialized morphological variants form a minority. Such a distribution is consistent with expectations from hierarchical galaxy formation models, where environmental processes and merger events progressively transform morphological characteristics over cosmic time.

\begin{figure}[h]
     % \plotone{quan.pdf}
     % \raisebox{-5pt}
     {\includegraphics[width=\columnwidth]{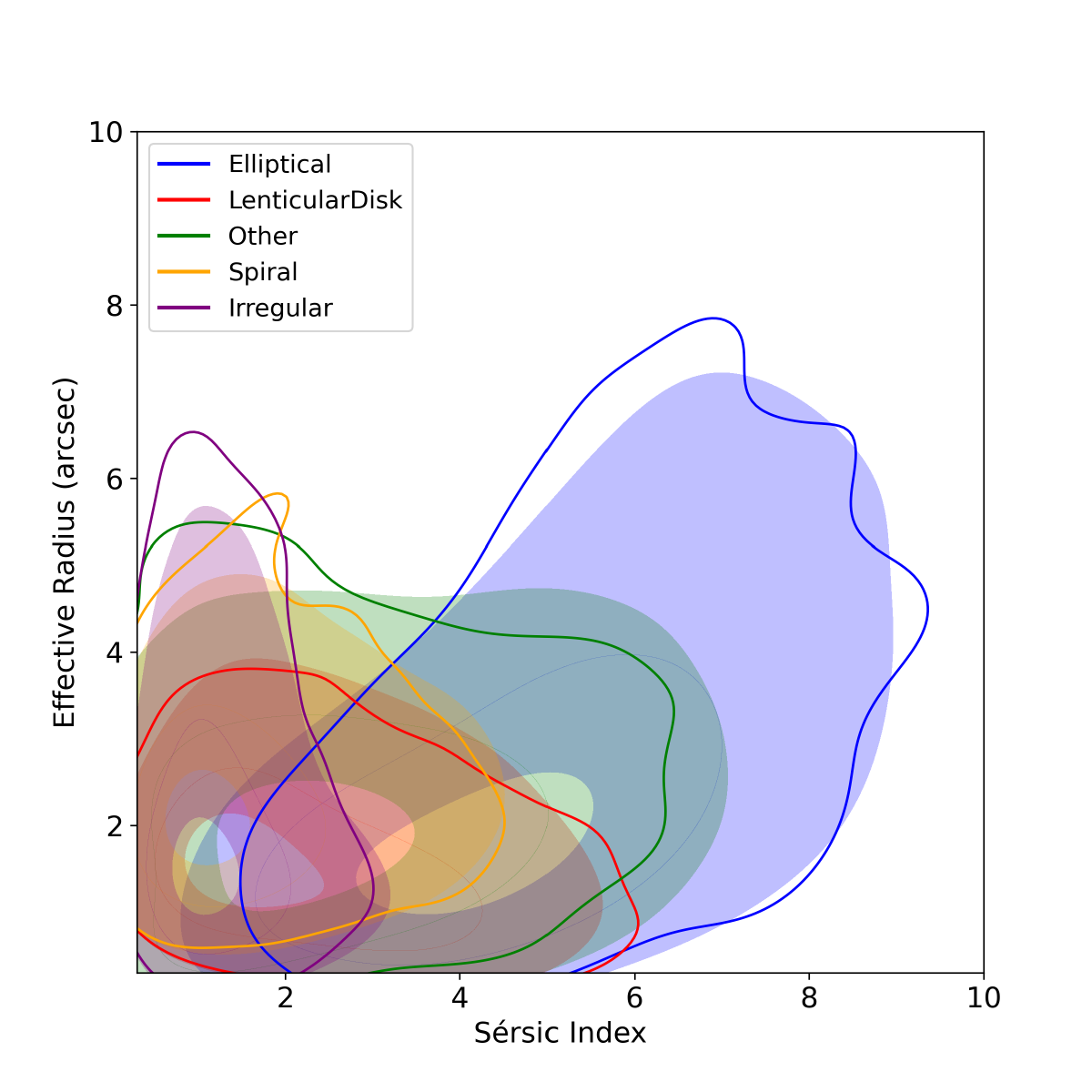}}
    % \caption{2D kernel density distributions of \textcolor{red}{the selected 310,000-galaxies} in the Sérsic index $n$ vs. effective radius $r_e$. The contours represent five morphological classes: Elliptical (blue), Lenticular-Disk (red), Other (green), Spiral (orange), and Irregular (purple). The contours, from innermost to outermost, represent the 16\%, 50\%, and 84\% cumulative probability regions for each morphological class, based on 2D kernel density estimation.}
    \caption{
Two-dimensional kernel density distributions in the Sérsic index $n$ vs. effective radius $r_e$ plane for the 305,583 classified galaxies (shaded regions) and the 5,000 labeled galaxies (contours). Shaded regions represent the 16\%, 50\%, and 84\% cumulative probability levels for each morphological class. For the labeled sample, only the 84\% probability contour is plotted for clarity. Morphological classes are color-coded as follows: Elliptical (blue), Lenticular–Disk (red), Other (green), Spiral (yellow), and Irregular (purple).
}
    \label{fig:r_n} 
\end{figure}

To further characterize these morphological classes, we examined their distribution in the parameter space of effective radius ($r_e$) as a function of S\'{e}rsic index ($n$), as presented in Figure \ref{fig:r_n}. This parameter space reveals distinctive clustering patterns corresponding to the fundamental physical properties of each galaxy type. Elliptical galaxies predominantly exhibit high S\'{e}rsic indices ($n\gtrsim2.5$), consistent with their concentrated light profiles and centrally dominated stellar distributions. Conversely, spiral and irregular galaxies display characteristically low S\'{e}rsic indices ($n\lesssim2.5$), reflecting their more extended, disk-dominated light distributions with exponential surface brightness profiles. Lenticular-disk galaxies occupy an intermediate parameter space with moderate S\'{e}rsic indices, consistent with their hybrid nature combining both bulge and disk components. The ``other" category, as expected, exhibits the most dispersed distribution across this parameter space, reflecting both its inherent morphological heterogeneity and the potential inclusion of some misclassified objects. In the same figure, a comparison with the $r_e-n$ density distributions of the 5,000 labeled galaxies showed excellent agreement, thereby validating the statistical consistency of our SSL-based classification.

A further visual examination of the 23,268 images classified as ``other" reveals a diverse collection of objects, including underexposed images, ring galaxies, shell galaxies, ongoing mergers, galaxy clusters, stellar contaminants, quasars, and gravitational lenses.
% , as illustrated in Figure \ref{fig:find}. 
It is important to note that while the model was not explicitly trained on each specific subclass within this ``other" category, the labeled training and testing datasets were consistently curated based on expert visual inspection. This rigorous labeling ensured that the model effectively learned to recognize morphologically distinct objects that deviate from the principal Hubble sequence categories. This capability highlights a significant potential for transfer learning applications, enabling our pre-trained representations to be fine-tuned for specialized astronomical objects such as gravitational lenses, ultra-diffuse galaxies, and merger systems.

% \vspace{1cm}
\section{Discussion} \label{sec:discussion}
In this section, we analyze the classification results, compare our method with traditional supervised learning, unsupervised approaches, and existing self-supervised frameworks. 
% In addition, we discuss the different settings in the training process and the effectiveness of our self-supervised learning approach.
We also highlight the broader implications of our findings for future astronomical surveys.

\subsection{Comparison with Other Machine Learning Methods} \label{sec:compare}
A rich literature has explored galaxy morphology classification under supervised, unsupervised, and self-supervised paradigms. Here we contextualize our approach with respect to representative methods in each category.

Among supervised approaches, \cite{rotation} reports near-perfect performance (accuracy exceeding 99\%) on Galaxy Zoo using a rotation-invariant convolutional network trained with high-quality labels. Their training set comprises 61{,}578 labeled images (with 10\% held for internal validation), and the evaluation dataset includes 79{,}975 images. This result sets a strong upper bound on task-specific accuracy in supervised learning. Although our absolute accuracy is lower than that of the nearly saturated supervised benchmark, our method pursues a different optimum: it achieves competitive performance with far fewer labels, relying on only 5{,}000 annotated images. This comparison underscores the efficiency of the label as a primary advantage of our framework.

Unsupervised methods eliminate the need for labels. For instance, \cite{hybird} achieve 95\% accuracy in separating spiral from elliptical galaxies using 36{,}000 SDSS DR7 images, with engineered EGG metrics that correlate strongly with physical properties and thus enable effective clustering. Our method targets a more demanding five-class taxonomy rather than a binary separation and dispenses with hand-crafted features. Although the resulting accuracy is somewhat lower than the binary unsupervised result, the learned representation is more general and better aligned with the complexity of galaxy morphology, favoring transferability across diverse structures instead of relying on task-specific descriptors.

Within self-supervised learning, recent studies combine modern architectures with contrastive objectives. \cite{selfsuper1} integrate vision transformers with CNN backbones and report 94.7\% and 96.5\% accuracy on three-class GZ2 and SDSS-DR17 tasks, respectively, and 89.9\% on a four-class GZ DECaLS task, using 7{,}168, 9{,}914, and 16{,}284 images. \cite{selfsuper2} propose the MCL-Galaxy model based on momentum contrastive learning, obtaining 90.12\% accuracy on a five-class classification task with 28{,}793 filtered labeled images. These results confirm that self-supervised pretraining can yield strong performance, but they still rely on a relatively large number of annotations. In contrast, our model uses only 5{,}000 labeled images together with 305{,}583 unlabeled images, learning robust and generalizable features for a five-class setting while markedly reducing the annotation burden. This evidences a favorable accuracy-per-label trade-off and demonstrates that scaling the unlabeled corpus in self-supervised training can substantially relax the need for large labeled datasets without sacrificing reasonable classification performance.

Taken together, comparisons across supervised, unsupervised, and self-supervised regimes highlight complementary strengths. Fully supervised methods can saturate accuracy when labels are plentiful; unsupervised pipelines excel at coarse separation with engineered features; and self-supervised strategies offer a middle ground. Our contribution is to push this middle ground toward label efficiency and representational generality: by leveraging abundant unlabeled data and avoiding hand-crafted features, we achieve competitive accuracy in a more granular five-class task with an order of magnitude fewer labels than many recent baselines.

\subsection{Implications for upcoming large surveys}
One of the main practical implications of our two-stage self-supervised framework is its suitability for the data volume of upcoming large imaging surveys. Modern surveys are producing galaxy images at scales that make traditional morphology pipelines increasingly difficult to sustain. Manual visual inspection is prohibitively expensive, while fully supervised deep learning approaches require large, carefully labeled samples that are often unavailable in current survey operations. In this context, a framework that can learn from abundant unlabeled images while requiring only a relatively small labeled subset for downstream classification is particularly valuable. Our results on \textit{KiDS} DR5 demonstrate that this strategy delivers robust morphology classification performance while substantially reducing the annotation burden, a central requirement for future survey-based galaxy studies.

A key scaling advantage of our approach is that the most expensive part of learning is moved from label collection to representation learning on unlabeled data. In our framework, the self-supervised pretraining stage is a one-time investment: it learns a general-purpose encoder for galaxy images. After the encoder captures useful morphological features, the downstream classification step becomes much lighter, requiring only a small labeled sample to train a simple classifier. This changes the task from repeatedly training large supervised models for each new classification problem to reusing a pretrained representation and adapting it with limited additional supervision. For surveys with abundant unlabeled images but costly expert classifications, this is a substantial benefit. It also enables morphology classification to reach larger samples than would be feasible with a purely supervised workflow. The framework is also advantageous from both computational and data-handling perspectives. Contrastive pretraining is well suited to modern parallel hardware, and it can be scaled using distributed training when larger datasets are available. In addition, the encoder compresses each image into a lower-dimensional feature representation, which can be useful beyond the immediate classification task. Although we focus here on morphology classification, such representations may also support related analyses, including similarity-based retrieval, clustering, and the identification of unusual systems, provided that these downstream applications are explicitly validated. In this sense, the self-supervised encoder serves not only as a classifier backbone, but also as a compact representation model for large survey image collections. For next-generation surveys, where storage and repeated image-level processing become increasingly costly, representation learning of this kind can provide an effective intermediate step between raw imaging data and science-driven analysis.

While these properties make our framework relevant for forthcoming surveys such as Euclid and CSST, it is important not to assume that the learned representation will transfer perfectly between surveys. A model trained on one survey (e.g., KiDS images) is not guaranteed to perform optimally when applied to data from another survey. Several forms of observational differences can affect performance. First, the apparent morphology depends on the effective bandpass and the image processing pipeline. Our inputs are based on $gri$ color-composite images from KiDS, whereas other surveys use different filter sets, which can change the relative visibility of star-forming clumps, dust lanes, and bulge–disk contrast. Second, differences in PSF, depth, noise properties, and spatial resolution may reduce or distort small-scale structures—such as weak spiral arms, bars, and asymmetries—that are important for classification. Third, our preprocessing uses a fixed $64 \times 64$ central crop, which corresponds to a specific angular scale in KiDS. The same pixel count in another survey corresponds to a different physical region of the galaxy and therefore captures a different fraction of its morphology. Nevertheless, the framework can still be adapted to other surveys. A realistic strategy is not to apply the model directly without retraining, but to adapt within the same two-stage framework. First, the encoder can be further trained on unlabeled images from the target survey, so that the learned representation matches its bandpass, PSF, resolution, and noise properties. Then, a small labeled subset from that survey can be used to train the classifier, or to perform limited supervised adaptation if needed. Therefore, the main value of our method is not to claim immediate cross-survey portability, but to provide a scalable, label-efficient foundation on which survey-specific morphology classifiers can be built and updated as new data become available.

\section{Conclusions}\label{sec:conslusion}
In this work, we confronted the growing challenge of galaxy morphology classification in an era of large-scale astronomical surveys. We have shown that traditional methods, such as manual classification, fully supervised learning, and classic unsupervised learning, each face significant limitations, whether in feasibility, the need for costly labels, or performance. This paper introduced a two-stage, self-supervised learning framework designed to resolve this problem. By first pretraining a neural network on unlabeled data, our model learns the rich, intrinsic features of galaxy morphology. This allows for a subsequent, highly efficient supervised step that achieves an accuracy of \textbf{$90.5\%\pm0.2\%$} with only a small number of labeled examples. Our method effectively combines the low-labeling cost of unsupervised methods with the high accuracy of supervised ones.

To validate its practical value, we successfully applied this framework to the Kilo-Degree Survey (KiDS) DR5 dataset. The result is a public morphology catalog of 310,583 galaxies,
% over 310k galaxies, 
one of the largest of its kind generated through automated deep learning. This achievement serves as a robust proof-of-concept, demonstrating that our method is not only accurate but also scalable and ready for production-level use on current survey data.

Looking forward, the efficiency and scalability of our self-supervised approach make it an ideal tool for upcoming surveys like CSST and Euclid, which will produce even larger volumes of data. This work confirms that self-supervised learning is a powerful and essential strategy for tackling the analytical challenges of modern astronomy, enabling us to fully exploit the scientific potential of these next-generation datasets.

\begin{acknowledgments}
This work is supported by the National Natural Science Foundation of China (Grant No. 12588202) and the China Manned Space Program with grant No. CMS-CSST-2025-A03. Rui Li acknowledges the National Natural Science Foundation of China (No. 12203050) and the Natural Science Foundation of Henan Province of China (Grant No. 252300423008). 
% Based on data obtained from the ESO Science Archive Facility with DOI: https://doi.org/10.18727/archive/37, and https://doi.eso.org/10.18727/archive/59 and on data products produced by the KiDS consortium.
This work is based on data obtained from the ESO Science Archive Facility \citep{eso_archive_37, eso_archive_59}, and on data products produced by the KiDS consortium.
The KiDS production team acknowledges support from: Deutsche Forschungsgemeinschaft, ERC, NOVA and NWO-M grants; Target; the University of Padova, and the University Federico II (Naples).
The authors acknowledge the use of ChatGPT and Gemini for assistance with English language editing and manuscript refinement.
\end{acknowledgments}

% \begin{contribution}

% All authors contributed equally to the Terra Mater collaboration.

% %
% \end{contribution}

% \appendix

\bibliographystyle{aasjournalv7}
% \bibliography{references}
\bibliography{sample701}

@ARTICLE{2025ApJS..279...26F,
       author = {{Feng}, Hai-Cheng and {Li}, Rui and {Napolitano}, Nicola R. and {Li}, Sha-Sha and {Bai}, J.~M. and {Dong}, Yue and {Li}, Ran and {Liu}, H.~T. and {Lu}, Kai-Xing and {Pan}, Zhi-Wei and {Radovich}, Mario and {Shan}, Huan-Yuan and {Wang}, Jian-Guo and {Xi}, Wen-Zhe and {Xie}, Ling-Hua and {Yuan}, Zun-Li and {Zhang}, Yang-Wei},
        title = "{Morpho-photometric Classification of KiDS DR5 Sources Based on Neural Networks: A Comprehensive Star-Quasar-Galaxy Catalog}",
      journal = {\apjs},
         year = 2025,
        month = jul,
       volume = {279},
       number = {1},
          eid = {26},
        pages = {26},
          doi = {10.3847/1538-4365/adde5a},
archivePrefix = {arXiv},
       eprint = {2406.03797},
 primaryClass = {astro-ph.GA},
       adsurl = {https://ui.adsabs.harvard.edu/abs/2025ApJS..279...26F}
}

@ARTICLE{2026ApJS..283...49L,
       author = {{Lv}, Jiameng and {Li}, Xu and {Cao}, Liang and {Gao}, Xi and {Li}, Nan and {Fu}, Mingxiang and {Li}, Yushan and {Duan}, Manni and {Jia}, Peng},
        title = "{FAMA{\textemdash}A Scalable Foundational Astronomical Masked Autoencoder for Astronomical Image Analysis}",
      journal = {\apjs},
         year = 2026,
        month = apr,
       volume = {283},
       number = {2},
          eid = {49},
        pages = {49},
          doi = {10.3847/1538-4365/ae4509},
       adsurl = {https://ui.adsabs.harvard.edu/abs/2026ApJS..283...49L}
}

@article{ellptical2006,
    author = {Renzini, Alvio},
    title = {Stellar Population Diagnostics of Elliptical Galaxy Formation},
    journal = {ARA\&A},
    volume = {44},
    pages = {141-192},
    year = 2006,
    doi = {10.1146/annurev.astro.44.051905.092450}
}

@article{spiral1998,
    author = {Kennicutt, Robert C., Jr.},
    title = {Star Formation in Galaxies Along the Hubble Sequence},
    journal = {ARA\&A},
    volume = {36},
    pages = {189-232},
    year = 1998,
    doi = {10.1146/annurev.astro.36.1.189}
}

@article{dwarf2009,
    author = {Tolstoy, E. and Hill, V. and Tosi, M.},
    title = {Star Formation Histories, Abundances and Kinematics of Dwarf Galaxies in the Local Group},
    journal = {ARA\&A},
    volume = {47},
    pages = {371-425},
    year = 2009,
    doi = {10.1146/annurev-astro-082708-101650}
}

@article{conselice,
    author = {Conselice, Christopher J},
    title = {The Evolution of Galaxy Structure Over Cosmic Time},
    journal = {ARA\&A},
    volume = {52},
    pages = {291-337},
    year = 2014,
    doi = {10.1146/annurev-astro-081913-040037},
    archivePrefix = {arXiv},
    eprint = {1403.2783},
}

@article{Krywult,
    author = {Krywult, J. and Tasca, L. A. M. and Pollo, A. and Vergani, D. and Bolzonella, M. and Davidzon, I. and Iovino, A. and Gargiulo, A. and Haines, C. P. and Scodeggio, M. and Guzzo, L. and Zamorani, G. and Garilli, B. and Granett, B. R. and de la Torre, S. and Abbas, U. and Adami, C. and Bottini, D. and Cappi, A. and Cucciati, O. and Franzetti, P. and Fritz, A. and Le Brun, V. and Le Fèvre, O. and Maccagni, D. and Małek, K. and Marulli, F. and Polletta, M. and Tojeiro, R. and Zanichelli, A. and Arnouts, S. and Bel, J. and Branchini, E. and Coupon, J. and De Lucia, G. and Ilbert, O. and McCracken, H. J. and Moscardini, L. and Takeuchi, T. T.},
    title = {The VIMOS Public Extragalactic Redshift Survey (VIPERS). The coevolution of galaxy morphology and colour to z 1},
    journal = {A\&A},
    pages = {A120},
    volume = {598},
    year = 2017,
    doi = {10.1051/0004-6361/201628953},
}

@article{Dressler1980,
    author = {Dressler, A.},
    title = {Galaxy morphology in rich clusters: implications for the formation and evolution of galaxies.},
    journal = {ApJ},
    pages = {351-365},
    volume = {236},
    year = 1980,
    doi = {10.1086/157753},
}

@article{Huerytas2024,
    author = {Huertas-Company, M. and Iyer, K. G. and Angeloudi, E. and Bagley, M. B. and Finkelstein, S. L. and Kartaltepe, J. and McGrath, E. J. and Sarmiento, R. and Vega-Ferrero, J. and Arrabal Haro, P. and Behroozi, P. and Buitrago, F. and Cheng, Y. and Costantin, L. and Dekel, A. and Dickinson, M. and Elbaz, D. and Grogin, N. A. and Hathi, N. P. and Holwerda, B. W. and others},
    title = {Galaxy morphology from $z \sim 6$ through the lens of JWST},
    journal = {A\&A},
    volume = {685},
    pages = {A48},
    year = 2024,
    doi = {10.1051/0004-6361/202346800},
}

@article{Huertas2016,
    author = {Huertas-Company, M. and Bernardi, M. and P{\'e}rez-Gonz{\'a}lez, P. G. and Ashby, M. L. N. and Barro, G. and Conselice, C. and Daddi, E. and Dekel, A. and Dimauro, P. and Faber, S. M. and Grogin, N. A. and Kartaltepe, J. S. and Kocevski, D. D. and Koekemoer, A. M. and Koo, D. C. and Mei, S. and Shankar, F.},
    title = {Mass assembly and morphological transformations since $z \sim 3$
from CANDELS},
    journal = {MNRAS},
    volume = {462},
    pages = {4495-4516},
    year = 2016,
    doi = {10.1093/mnras/stw1866}
}

@book{Conselice2020,
    author = {Conselice, Christopher J.},
    title = {The Cosmic Evolution of Galaxy Structure},
    publisher = {IOP Publishing},
    year = 2020,
    doi = {10.1088/2514-3433/abb602},
}

@book{Holwerda2021,
  author = {Holwerda, Benne W.},
  title = {Galaxy Morphology},
  publisher = {IOP Publishing},
  year = {2021},
  doi = {10.1088/2514-3433/ac2c7d}
}

@article{Roy2018Sersic,
    author = {Roy, N. and Napolitano, N. R. and La Barbera, F. and
       Tortora, C. and Getman, F. and Radovich, M. and
       Capaccioli, M. and Brescia, M. and Cavuoti, S. and
       Longo, G. and Raj, M. A. and Puddu, E. and Covone, G. and
       Amaro, V. and Vellucci, C. and Grado, A. and Kuijken, K. and
       Verdoes Kleijn, G. and Valentijn, E.},
    title = {Evolution of galaxy size-stellar mass relation from the Kilo-Degree Survey},
    journal = {MNRAS},
    volume = {480},
    pages = {1057--1080},
    year = 2018,
    doi = {10.1093/mnras/sty1917},
}

@article{CAS,
    author = {Christopher J. Conselice},
    title = {The Relationship between Stellar Light Distributions of Galaxies and Their Formation Histories},
    journal = {ApJS},
    volume = {147},
    pages = {1-28},
    year = 2003,
    doi = {10.1086/375001},
}

@book{Sersic,
    author = {S{\'e}rsic, Jose Luis},
    year      = {1968},
  title     = {Atlas de Galaxias Australes},
  publisher = {Observatorio Astron{\'o}mico},
  address   = {Cordoba, Argentina}
}

@article{CAScite2024,
    author = {Chantavat, T.  and Yuma, S.  and  Malelohit, P. and others},
    title = {Morphological Evolution of Disk Galaxies and Their Concentration, Asymmetry, and Clumpiness (CAS) Properties in Simulations across Toomre's Q Parameter},
    journal = {ApJ},
    volume = {965},
    pages = {77},
    year = 2024,
    doi = {10.3847/1538-4357/ad3218},
}

@article{CAScitejwst2023,
    author = {Kartaltepe, Jeyhan S. and Rose, Caitlin and Vanderhoof, Brittany N. and McGrath, Elizabeth J. and Costantin, Luca and others},
    title = {CEERS Key Paper. III. The Diversity of Galaxy Structure and Morphology at z = 3-9 with JWST},
    journal = {ApJL},
    volume = {946},
    pages = {L15},
    year = 2023,
    doi = {10.3847/2041-8213/acad01},
}

@article{Sersic2019,
    author = {Breda, Iris and Papaderos, Polychronis and  Gomes, Jean Michel and Amarantidis, Stergios},
    title = {A new fitting concept for the robust determination of Sérsic model parameters},
    journal = {A\&A},
    volume = {632},
    pages = {A128},
    year = 2019,
    doi = {10.1051/0004-6361/201935144},
}

@article{neareast2024,
    author = {Mukundan, Kavya and Nair, Preethi  and Bailin, Jeremy and others},
    title = {Automating galaxy morphology classification using k-nearest neighbours and non-parametric statistics},
    journal = {MNRAS},
    volume = {533},
    pages = {292-312},
    year = 2024,
    doi = {10.1093/mnras/stae1684},
}

@article{dense4_2025,
    author = {Mao, Yu and Tu, Liangping and  Xu, Zhenyang and  Jiang, Yue and  Zheng, Mingyu},
    title = {Galaxy Morphology Classification Based on DenseNet-SE4 Algorithm},
    journal = {Research in Astronomy and Astrophysics},
    volume = {25},
    pages = {085010},
    year = 2025,
    doi = {10.1088/1674-4527/ade22c},
}

@article{HSTfit2011,
    author = {Hoyos, Carlos and den Brok, Mark and Verdoes Kleijn, Gijs and Carter, David and Balcells, Marc and Guzm{\'a}n, Rafael and Peletier, Reynier and Ferguson, Henry C. and Goudfrooij, Paul and Graham, Alister W. and Hammer, Derek and Karick, Arna M. and Lucey, John R. and Matkovi{\'c}, Ana and Merritt, David and Mouhcine, Mustapha and Valentijn, Edwin},
    title = {The HST/ACS Coma Cluster Survey - III. Structural parameters of galaxies using single Sérsic fits},
    journal = {MNRAS},
    volume = {411},
    pages = {2439-2460},
    year = {2011},
    doi = {10.1111/j.1365-2966.2010.17855.x},
}

@article{JWST2025,
    author = {Genin, Aur{\'e}lien and Shuntov, Marko and Brammer, Gabe and Allen, Natalie and Ito, Kei and Magdis, Georgios and Matharu, Jasleen and Oesch, Pascal A. and Toft, Sune and Valentino, Francesco},
    title = {DAWN JWST Archive: Morphology from profile fitting of over 340 000 galaxies in major JWST fields: Morphology evolution with redshift and galaxy type},
    journal = {A\&A},
    volume = {699},
    pages = {A343},
    year = 2025,
    doi = {10.1051/0004-6361/202555504},
}

@article{Euclid2025,
    author = {{Euclid Collaboration} and Quilley, L. and Damjanov, I. and others},
    title = {Euclid Quick Data Release (Q1). Exploring galaxy morphology across cosmic time through S{'e}rsic fits},
    journal = {arXiv e-prints},
    year = 2025,
    archivePrefix = {arXiv},
    eprint = {2503.15309},
    doi = {10.48550/arXiv.2503.15309}
}

@article{KiDSfit2025,
    author = {Georgiou, Christos and Chisari, Nora Elisa and Bilicki, Maciej and La Barbera, Francesco and Napolitano, Nicola R. and Roy, Nivya and Tortora, Crescenzo},
    title = {Intrinsic galaxy alignments in the KiDS-1000 bright sample: Dependence on colour, luminosity, morphology, and galaxy scale},
    journal = {A\&A},
    volume = {699},
    pages = {A252},
    year = 2025,
    doi = {10.1051/0004-6361/202554134},
}

@article{parmwrong2011,
    author = {Andrae, René and Jahnke, Knud  and  Melchior, Peter},
    title = {Parametrizing arbitrary galaxy morphologies: potentials and pitfalls},
    journal = {MNRAS},
    volume = {411},
    pages = {385-401},
    year = 2011,
    doi = {10.1111/j.1365-2966.2010.17690.x},
}

@article{Sericspiral2022,
    author = {Sonnenfeld, Alessandro},
    title = {The effect of spiral arms on the Sérsic photometry of galaxies},
    journal = {A\&A},
    volume = {659},
    pages = {A141},
    year = 2022,
    doi = {10.1051/0004-6361/202142786},
}

@article{Sen,
    author = {Vika, Marina and Vulcani, Benedetta and Bamford, Steven P. and Häußler, Boris and Rojas, Alex L.},
    title = {MegaMorph: classifying galaxy morphology
using multi-wavelength Sérsic profile fits},
    journal = {A\&A},
    volume = {577},
    page = {A97},
    year = 2015,
    doi = {10.1051/0004-6361/201425174},
}

@article{DES,
    author = {{The Dark Energy Survey Collaboration}},
    title = {The Dark Energy Survey},
    journal       = {arXiv e-prints},
    eprint        = {astro-ph/0510346},
    archivePrefix = {arXiv},
    year          = {2005},
    doi = {10.48550/arXiv.astro-ph/0510346},
}

@article{KiDS2013,
    author = {de Jong, Jelte T. A. and  Verdoes Kleijn, Gijs A. and  Kuijken, Konrad H. and  Valentijn, Edwin A.},
    title = {The Kilo-Degree Survey},
    journal = {Exp. Astron.},
    volume = {35},
    page = {25-44},
    year = 2013,
    doi = {10.1007/s10686-012-9306-1},
}

@article{LSST,
    author = {Ivezi\'c, \v{Z}eljko and Kahn, Steven M. and Tyson, J. Anthony and others},
    title = {LSST: From Science Drivers to Reference Design and Anticipated Data Products},
    journal = {ApJ},
    volume = {873},
    page = {111},
    year = 2019,
    doi = {10.3847/1538-4357/ab042c},
}

@article{GalaxyZoo,
    author = {Lintott, Chris J. and  Schawinski, Kevin and Slosar, Anže and  Land, Kate and Bamford, Steven and  Thomas, Daniel and  Raddick, M. Jordan and Nichol, Robert C. and  Szalay, Alex and  Andreescu, Dan and  Murray, Phil and Vandenberg, Jan},
    title = {Galaxy Zoo: morphologies derived from visual inspection of galaxies from the Sloan Digital Sky Survey},
    journal = {MNRAS},
    year = 2008,
    volume = 389,
    pages = {1179-1189},
    doi = {10.1111/j.1365-2966.2008.13689.x},
}

@article{GalaxyZoo1,
    author = {Chris Lintott and Kevin Schawinski and Steven Bamford and Anže Slosar and Kate Land and, Daniel Thomas and  Edd Edmondson and  Karen Masters and  Robert C. Nichol and  M. Jordan Raddick and et.al},
    title = {Galaxy Zoo 1: data release of morphological classifications for nearly 900 000 galaxies},
    journal = {MNRAS},
    year = 2011,
    volume = {410},
    pages = {166-178},
    doi = {10.1111/j.1365-2966.2010.17432.x},
}

@article{SDSS,
    author = {York, Donald G. and Adelman, J. and Anderson, John E., Jr. and others},
    title = {THE SLOAN DIGITAL SKY SURVEY : TECHNICAL SUMMARY},
    journal = {AJ},
    year = 2000,
    volume = 120,
    pages = {1579-1587},
    doi = {10.1086/301513},
}

@article{rotation,
    author = {Sander Dieleman and Kyle W. Willett and Joni Dambre1},
    title = {Rotation-invariant convolutional neural networks for galaxy morphology prediction},
    journal = {MNRAS},
    year = 2015,
    volume = {450},
    pages = {1441-1459},
    doi = {10.1093/mnras/stv632},
}

@article{cnn1,
    author = {Zhu, Xiao-Pan and Dai, Jia-Ming  and Bian, Chun-Jiang and Chen, Yu and Chen, Shi and Hu, Chen},
    title = {Galaxy morphology classification with deep convolutional neural
networks},
    journal = {Astrophys. Space Sci.},
    volume = {364},
    pages = {55},
    year = 2019,
    doi = {10.1007/s10509-019-3540-1},
}

@article{cnn2,
    author = {Kalvankar, Shreyas and Pandit, Hrushikesh and Parwate, Pranav},
    title         = {Galaxy Morphology Classification using EfficientNet Architectures},
    journal       = {arXiv e-prints},
    year          = {2020},
    eprint        = {2008.13611},
    archivePrefix = {arXiv},
    primaryClass  = {cs.CV},
}

@article{gini,
    author = {Abraham, Roberto G. and  van den Bergh, Sidney and  Nair, Preethi},
    title = {A New Approach to Galaxy Morphology. I. Analysis of the Sloan Digital Sky Survey Early Data Release},
    journal = {ApJ},
    volume = {588},
    pages = {218-229},
    year = 2003,
    doi = {10.1086/373919},
}

@article{gini2004,
    author = {Lotz, Jennifer M. and Primack, Joel and Madau, Piero},
    title = {A New Nonparametric Approach to Galaxy Morphological Classification},
    journal = {AJ},
    year = 2004,
    volume = {128},
    pages = {163-182},
    doi = {10.1086/421849},
}

@article{sersic2005,
    author = {Graham, Alister W. and Driver, Simon P. and Petrosian, Vahé and Conselice, Christopher J. and Bershady, Matthew A. and Crawford, Steven M. and Goto, Tomotsugu},
    title = {Total Galaxy Magnitudes and Effective Radii from Petrosian Magnitudes and Radii},
    journal = {AJ},
    year = 2005,
    volume = {130},
    pages = {1535-1544},
    doi = {10.1086/444475},
}

@article{gini2016,
    author = {Florian, Michael K.  and Li, Nan  and Gladders, Michael D.},
    title = {The Gini Coefficient as a Morphological Measurement of Strongly Lensed Galaxies in the Image Plane},
    journal = {ApJ},
    volume = {832},
    pages = {168},
    year = 2016,
    doi = {10.3847/0004-637X/832/2/168},
}

@article{asym,
    author = {Conselice, Christopher J. and Bershady, Matthew A. and Jangren, Anna},
    title = {The Asymmetry of Galaxies: Physical Morphology for Nearby and High-Redshift Galaxies},
    journal = {ApJ},
    volume = {529},
    pages = {886--910},
    year = 2000,
    doi = {10.1086/308300},
}

@article{Beys,
    author = {Serrano-Pérez, Jonathan and Díaz Hernández, Raquel and Sucar, L. Enrique},
    title = {Bayesian and convolutional networks for hierarchical morphological classification of galaxies},
    journal = {Exp. Astron.},
    volume = {58},
    number = {2},
    pages = {Article ID 5},
    year = 2024,
    doi = {10.1007/s10686-024-09950-y},
}

@article{pre,
    author = {Schneider, Jesse and Stenning, David C. and Elliott, Lloyd T.},
    title = {Efficient galaxy classification through pretraining},
    journal = {Front. Astron. Space Sci.},
    volume = {10},
    year = 2023,
    doi = {10.3389/fspas.2023.1197358},
}

@article{coder,
    author = {Spindler, Ashley and Geach, James E. and Smith, Michael J.},
    title = {AstroVaDEr: astronomical variational deep embedder for unsupervised
morphological classification of galaxies and synthetic image generation},
    journal = {MNRAS},
    volume = {502},
    pages = {985-1007},
    year = 2021,
    doi = {10.1093/mnras/staa3670},
}

@article{autoencoder0,
    author = {Fielding, Ezra and Nyirenda, Clement N. and Vaccari, Mattia},
    title = {The Classification of Optical Galaxy Morphology Using Unsupervised Learning Techniques},
    journal = {Proc. ICECET 2022},
    publisher = {IEEE},
    pages = {1-6},
    year = 2022,
    doi = {10.1109/ICECET55527.2022.9872611},
}

@article{autoencoder,
    author = {Seo, Eunsuk and Kim, Suk and Lee, Youngdae and Han, Sang-Il and Kim, Hak-Sub and Rey, Soo-Chang and Song, Hyunmi},
    title = {Similar Image Retrieval using Autoencoder. I. Automatic Morphology
Classification of Galaxies},
    journal = {PASP},
    volume = {135},
    number = {1050},
    year = {2023},
    
    pages = {17},
    doi = {10.1088/1538-3873/ace851},
}

@article{hybird,
    author = {Kolesnikov, I. and Sampaio, V. M. and  de Carvalho, R. R. and Conselice, C. and Rembold, S. B. and Mendes, C. L. and Rosa, R. R.},
    title = {Unveiling galaxy morphology through an unsupervised-supervised  hybrid approach},
    journal = {MNRAS},
    volume = {528},
    pages = {82-107},
    year = 2024,
    doi = {10.1093/mnras/stad3934},
}

@article{gini08,
    author = {Lisker, Thorsten},
    title = {IS THE GINI COEFFICIENT A STABLE MEASURE OF GALAXY STRUCTURE?},
    journal = {ApJS},
    volume = {179},
    pages = {319-325},
    year = 2008,
    doi = {10.1086/591795}
}

@article{selfsuper1,
    author = {Wei, Shoulin and  Li, Yadi and Lu, Wei and Li, Nan  and  Liang, Bo ;  Dai, Wei and Zhang, Zhijian},
    title = {Unsupervised Galaxy Morphological Visual Representation with Deep
Contrastive Learning},
    journal = {PASP},
    volume = {134},
    number = {1041},
    pages  = {id.114508},
    year = 2022,
    doi = {10.1088/1538-3873/aca04e},
}

@article{selfsuper2,
    author = {Shen, Guoqiang and Zou, Zhiqiang and Luo, A. -Li and Hong, Shuxin and Kong, Xiao},
    title = {A Galaxy Morphology Classification Model Based on Momentum Contrastive
Learning},
    journal = {PASP},
    volume = {135},
    number = {1052},
    pages = {104501},
    year = 2023,
    doi = {10.1088/1538-3873/acf8f7},
}

@article{densenet1,
    author = {Hui, Wuyu and Robert Jia, Zheng and Li, Hansheng and  Wang, Zijian},
    title = {Galaxy Morphology Classification with DenseNet},
    journal = {J. Phys. Conf. Ser.},
    volume = {2402},
    number = {1},
    pages = {012009},
    year = 2022,
    
    doi = {10.1088/1742-6596/2402/1/012009},
}

@article{densenet2,
    author = {Khramtsov, V. and Vavilova, I. B. and Dobrycheva, D. V. and Vasylenko, M. Yu. and Melnyk, O. V. and Elyiv, A. A. and Akhmetov, V. S. and Dmytrenko, A. M.},
    title = {Machine learning technique for morphological classification of galaxies from the SDSS. III. Image-based inference of detailed features},
     journal = {Space Sci. Technol.},
    volume  = {28},
    pages =  {27--55},
    number  = {5},
    year = 2022,
    doi = {10.15407/knit2022.05.027},
}

@article{densenet3,
    author = {Wang, Guangze},
    title = {Galaxy morphology classification with densenet},
    journal = {J. Phys. Conf. Ser.},
    volume = {2580},
    number = {1},
    pages = {012064},
    year = 2023,
    publisher = {IOP Publishing},
    doi = {10.1088/1742-6596/2580/1/012064},
}

@article{twoway,
    author = {Barchi, P. H. and de Carvalho, R. R. and Rosa, R. R. and Sautter, R. A. and Soares-Santos, M. and Marques, B. A. D. and Clua, E. and  Gonçalves, T. S. and de Sá-Freitas, C. and Moura, T. C.},
    title = {Machine and Deep Learning applied to galaxy morphology - A comparative study},
    journal = {Astron. Comput.},
    volume = {30},
    year = 2020,
    doi = {10.1016/j.ascom.2019.100334},
}

@article{imp,
    author = {Urechiatu, Raul and Frincu, Marc},
    title = {Improved Galaxy Morphology Classification with Convolutional Neural Networks},
    journal = {Universe},
    volume = {10},
    number = {6},
    year = 2024,
    article-id = {230},
    doi = {10.3390/universe10060230},
}

@article{simclr2020,
    author = {Ting Chen and Simon Kornblith and Mohammad Norouzi and Geoffrey Hinton},
    title = "{A Simple Framework for Contrastive Learning of Visual Representations}",
    journal = {arXiv:2002.05709},
    year = 2020,
    doi = {10.48550/arXiv.2002.05709},
archivePrefix = {arXiv},
}

@article{resnet_50,
    author = {KaiMing He and Xiangyu Zhang and Shaoqing Ren and Jian Sun},
    title = "{Deep Residual Learning for Image Recognition}",
    journal = {arXiv:1512.03385},
    year = 2015,
    doi = {10.48550/arXiv.1512.03385},
archivePrefix = {arXiv},  
}

@article{tem,
  author       = {Feng Wang and
                  Huaping Liu},
  title        = {Understanding the Behaviour of Contrastive Loss},
  journal       = {arXiv e-prints},
  year         = {2020},
  eprint       = {2012.09740},
  doi = {10.48550/arXiv.2012.09740},
}

@article{adam,
    author = {Kingma, Diederik P. and Ba, Jimmy},
    title = {Adam: A Method for Stochastic Optimization},
    journal       = {arXiv e-prints},
    eprint = {arXiv:1412.6980},
    year = 2014,
    doi = {10.48550/arXiv.1412.6980},
}

@article{hubble1926,
    author = {Hubble, E. P.},
    title = {Extragalactic nebulae},
    journal = {ApJ},
    volume = {64},
    pages = {321--369},
    year = 1926,
    doi = {10.1086/143018}
}

@article{KiDS5,
    author = {Wright, Angus H. and Kuijken, Konrad and Hildebrandt, Hendrik and et al.},
    title = {The fifth data release of the Kilo Degree Survey: Multi-epoch optical/NIR imaging covering wide and legacy-calibration fields},
    journal = {A\&A},
    volume = {686},
    pages = {A170},
    year = 2024,
    doi = {10.1051/0004-6361/202346730},
}

@article{Sextractor,
    author = {Bertin, E. and Arnouts, S.},
    title = {SExtractor: Software for source extraction.},
    journal = {A\&AS},
    volume = {117},
    pages = {393-404},
    year = 1996,
    doi = {10.1051/aas:1996164},
}

@article{GaZnet,
    author = {Li, Rui and Napolitano, Nicola R. and Feng, Haicheng and Li, Ran and Amaro, Valeria and Xie, Linghua and Tortora, Crescenzo and Bilicki, Maciej and Brescia, Massimo and Cavuoti, Stefano and Radovich, Mario},
    title = {Galaxy morphoto-Z with neural Networks (GaZNets)
I. Optimized accuracy and outlier fraction from imaging and photometry},
    journal = {A\&A},
    volume = {666},
    pages = {A85},
    year = 2022,
    doi = {10.1051/0004-6361/202244081},
}

@article{GaLnet,
    author = {Li, R. and Napolitano, N. R. and Roy, N. and Tortora, C.  and La Barbera, F. and Sonnenfeld, A. and Qiu, C. and Liu, S.},
    title = {Galaxy Light Profile Convolutional Neural Networks (GaLNets). I. Fast and Accurate Structural Parameters for Billion-galaxy Samples},
    journal = {ApJ},
    volume = {929},
    pages = {152},
    year = 2022,
    doi = {10.3847/1538-4357/ac5ea0},
}

@article{mass,
    author = {Xie, Linghua and  Napolitano, Nicola R. and Guo, Xiaotong  and  Tortora, Crescenzo and Feng, Haicheng  and Katsianis, Antonios and Li, Rui and  Wu, Sirui and  Radovich, Mario and Hunt, Leslie K. and  Wang, Yang  and Tang, Lin and Tang, Baitian and Huang, Zhiqi},
    title = {Toward a stellar population catalog in the Kilo Degree Survey: The impact of stellar recipes on stellar masses and star formation rates},
    journal = {Science China Physics, Mechanics \& Astronomy},
    volume = {66},
    pages = {129513},
    year = 2023,
    doi = {10.1007/s11433-023-2173-8},
}

@article{CSST,
    author = {Gong, Yan and Liu, Xiangkun and Cao, Ye and Chen, Xuelei and Fan, Zuhui and Li, Ran and Li, Xiao-Dong and Li, Zhigang and Zhang, Xin and Zhan, Hu},
    title = {Cosmology from the Chinese Space Station Optical Survey (CSS-OS)},
    journal = {ApJ},
    volume = {883},
    pages = {203},
    year = 2019,
    doi = {10.3847/1538-4357/ab391e},
}

@article{Euclid,
    author = {Laureijs, R. and Amiaux, J. and  Arduini, S. and Auguères, J. -L. and  Brinchmann, J. and Cole, R. and  Cropper, M. etc.},
    title         = {Euclid Definition Study Report},
  journal       = {arXiv e-prints},
  year          = {2011},
  eprint        = {1110.3193},
  archivePrefix = {arXiv},
  primaryClass  = {astro-ph.CO},
}

@misc{eso_archive_37,
  doi = {10.18727/ARCHIVE/37},
  url = {https://doi.eso.org/10.18727/archive/37},
  author = {Kuijken,  Konrad},
  language = {en},
  title = {KIDS - The Kilo-Degree Survey (VST)},
  publisher = {European Southern Observatory (ESO)},
  year = {2015},
  copyright = {Data Access Policy for ESO Data held in the ESO Science Archive Facility}
}

@misc{eso_archive_59,
  doi = {10.18727/ARCHIVE/59},
  url = {https://doi.eso.org/10.18727/archive/59},
  author = {Edge,  Alastair},
  language = {en},
  title = {VIKING - VISTA Kilo-degree Infrared Galaxy Survey},
  publisher = {European Southern Observatory (ESO)},
  year = {2013},
  copyright = {Data Access Policy for ESO Data held in the ESO Science Archive Facility}
}

@article{Hayat2021,
  title = {Self-supervised Representation Learning for Astronomical Images},
  volume = {911},
  ISSN = {2041-8213},
  url = {http://dx.doi.org/10.3847/2041-8213/abf2c7},
  DOI = {10.3847/2041-8213/abf2c7},
  number = {2},
  journal = {ApJL},
  publisher = {American Astronomical Society},
  author = {Hayat,  Md Abul and Stein,  George and Harrington,  Peter and Lukić,  Zarija and Mustafa,  Mustafa},
  year = {2021},
  month = apr,
  pages = {L33}
}

@article{BaronPerez2025,
  title = {Classification of radio sources through self-supervised learning},
  volume = {699},
  ISSN = {1432-0746},
  url = {http://dx.doi.org/10.1051/0004-6361/202554735},
  DOI = {10.1051/0004-6361/202554735},
  journal = {A\&A},
  publisher = {EDP Sciences},
  author = {Baron Perez,  Nicolas and Br\"{u}ggen,  Marcus and Kasieczka,  Gregor and Lucie-Smith,  Luisa},
  year = {2025},
  month = {july},
  pages = {A302}
}

@article{Yang2025,
  title = {Searching for Strong Lenses from DESI Legacy Surveys with a Hybrid CNN-Transformer Architecture with Self-supervised Learning},
  volume = {137},
  ISSN = {1538-3873},
  url = {http://dx.doi.org/10.1088/1538-3873/ade400},
  DOI = {10.1088/1538-3873/ade400},
  number = {6},
  journal = {PASP},
  publisher = {IOP Publishing},
  author = {Yang,  Jinrui and Li,  Nan and He,  Zizhao and Li,  Tian and Zou,  Zhiqiang and Shen,  Shiyin},
  year = {2025},
  month = {june},
  pages = {064504}
}

@article{Zuo2025,
  title = {FALCO: Foundation Model of Astronomical Light Curves for Time Domain Astronomy. Implementation and Applications on Kepler Data},
  volume = {171},
  ISSN = {1538-3881},
  url = {http://dx.doi.org/10.3847/1538-3881/ae1467},
  DOI = {10.3847/1538-3881/ae1467},
  number = {1},
  journal = {AJ},
  publisher = {American Astronomical Society},
  author = {Zuo,  Xiaoxiong and Tao,  Yihan and Huang,  Yang and Kang,  Zhixuan and Chen,  Huaxi and Cui,  Chenzhou and Pan,  Jiashu and Kong,  Xiao and Ting,  Yuan-Sen and Tang,  Xiaoyu and Han,  Henggeng and Mu,  Haiyang and Xu,  Yunfei and Fan,  Dongwei and Xue,  Guirong and Luo,  Ali and Liu,  Jifeng},
  year = {2025},
  month = Dec,
  pages = {10}
}

\end{document}